\documentclass[12pt]{article}
\usepackage{epsfig}
\usepackage{feynmp}

\newcommand {\sla}[1] {#1\!\!\! /}
\newcommand {\be} {\begin{equation}}
\newcommand {\ee} {\end{equation}}
\newcommand {\bes} {\begin{displaymath}}
\newcommand {\ees} {\end{displaymath}}
\newcommand {\ba} {\begin{eqnarray}}
\newcommand {\ea} {\end{eqnarray}}
\newcommand {\al} {\alpha}
\newcommand {\alo} {{\alpha_{0}}}
\newcommand {\as} {{\sf a}}

\newcommand {\bt} {\beta}
\newcommand {\bto} {{\beta_0}}

\newcommand {\eps} {\epsilon}
\newcommand {\no} {\nonumber}
\newcommand {\da} {\dagger}
\newcommand {\mi} {p_{-}}
\newcommand {\pl} {p_+}

\newcommand{\vk}{v\!\cdot\! k}

\def\slash#1{#1 \hskip -0.5em / }
\def\Pp{\frac{1 + \slash{v}}{2}}
\def\Pm{\frac{1 - \slash{v}}{2}}
\def\gl#1{(\ref{#1})}
\def\L#1{{\cal L}_{\mbox{\scriptsize #1}}}
\def\tr#1{{\rm tr}\left[#1\right]}

\begin{document}



\thispagestyle{empty}
\begin{titlepage}

\begin{flushright}
hep-ph/9601257 \\
\end{flushright}
\vspace{0.3cm}
\begin{center}
\Large \bf
Heavy baryons in the quark--diquark picture 
 \\
\end{center}
\vspace{0.5cm}
\renewcommand{\thefootnote}{\fnsymbol{footnote}}
\begin{center}
Dietmar Ebert\footnotemark[1],
Thorsten Feldmann\footnotemark[1]\footnotemark[4], \\
{\sl Institut f\"ur Physik, Humboldt--Universit\"at zu Berlin,\\
 Invalidenstra\ss e 110, D--10115 Berlin, Germany} \\
\end{center}
\vspace{0.3cm}
\begin{center}
Christiane Kettner\footnotemark[2],
Hugo Reinhardt\footnotemark[3],\\
{\sl Institut f\"ur Theoretische Physik,
      Universit\"at T\"ubingen,\\
   Auf der Morgenstelle 14, D--72076 T\"ubingen, Germany}
\end{center}
\vspace{0.6cm}
%
\begin{abstract}
\noindent
We describe heavy baryons as bound states of a quark and a diquark. 
For this purpose we derive the Faddeev equation for baryons containing
a single heavy quark from a Nambu--Jona-Lasinio type of model which is 
appropriately  extended to include also heavy quarks.
The latter are treated in the heavy mass limit. The 
heavy baryon Faddeev equation is then solved
using a static approximation for the exchanged quark.
\end{abstract}

%
\vspace{2em}
\begin{center}
({\em To appear in Int.\ Journ.\ Mod.\ Phys. {\bf A}})
\end{center}
\vspace{0.3cm}
\setcounter{footnote}{1}
\footnotetext{Supported by
{\it Deutsche Forschungsgemeinschaft} under contract Eb 139/1--2.}
\setcounter{footnote}{2}
\footnotetext{Supported by {\it Graduiertenkolleg Hadronen und Kerne}}
\setcounter{footnote}{3}
\footnotetext{Supported by {\it Deutsche Forschungsgemeinschaft}
under contract Re 856/2-2.}
\setcounter{footnote}{4}
\footnotetext{Address since 01/03/97: {\it Institut f\"ur
Physik, Universit\"at Wuppertal, D--42097 Wuppertal, Germay}\/.}
\vfill
\end{titlepage}

\renewcommand{\thefootnote}{\arabic{footnote}}
\setcounter{footnote}{0}
\setcounter{page}{1}


\section{Introduction}

The physics of hadrons with one heavy quark coupled to some 
light degrees of freedom becomes much simpler in the heavy quark limit
due to the arising heavy flavor and spin symmetries.
The basic concepts have been worked out within the past few years 
\cite{HQET} (see also \cite{hqet} for recent reviews).
The heavy quark symmetries are realized approximately
as long as the heavy quark mass $m_Q$ is much larger than a typical 
QCD scale, $m_Q \gg \Lambda_{QCD}$, which is the case for $Q= b,c$.
Phenomenologically, they result in an approximate degeneracy of heavy 
hadrons differing only in the heavy quark spin, and in a
definite scaling behavior of observables with the heavy quark masses.\\
Since the heavy quark inside the hadron has to be almost on--shell, it 
is useful to reformulate the heavy quark dynamics in terms of the 
off--shell momentum $k = p - m_Q v$, where  $p$ is the total momentum 
and $v$ the four--velocity ($v^2=1$) of the heavy quark, with 
$k^2 \ll m_Q^2$.
Furthermore the heavy quark spinor is split into large and small 
components by means of the projection operators $\Pp$ and $ \Pm$.
In the heavy mass limit ($m_Q \to \infty$) only the large component
\be
Q_v(x) = \Pp \, e^{ i \, m_Q \, v \cdot x} \, Q(x) \ , 
        \qquad  (\slash{v} Q_v = Q_v) \label{trafo}
\ee
survives, and the heavy quark part of the QCD Lagrangian  reads
\be
\L{\rm HQET} = \bar{Q}_v (i \, v \!\cdot\! {\cal D}) Q_v
        + O(1/m_Q) 
\  ,
\no
\ee
where ${\cal D}_\mu$ is the covariant derivative of the strong 
interaction.
Note that the heavy quark's velocity $v_\mu$ is a conserved quantity
in the infinite mass limit.  
The operator $(i v \!\cdot\! {\cal D})$ is obviously insensitive to 
spin and flavor quantum numbers of the heavy quark, therefore the 
corresponding hadron properties are determined by the light quarks 
only. 
The systematic expansion of QCD in powers of $1/m_Q$ 
results in the so--called 
Heavy Quark Effective Theory (HQET) \cite{MRR92,KT91}.
Recently, the ideas of HQET have been successfully extended to the 
sector with two heavy quarks (Heavy Quarkonium Effective Theory
\cite{HQQET}). 
Heavy fermion effective field theory techniques are also employed to 
derive a consistent expansion scheme for baryon chiral perturbation 
theory with $SU(3)_F$ flavor symmetry \cite{HFET}.\\
Although HQET provides the tool for calculating the perturbative QCD
corrections to short--distance processes, one is left with the 
non--perturbative features of QCD that lead to the formation of
hadrons (containing also light quarks) at low energies.
As is well known, the low energy properties of light quark systems 
are dictated by the approximate chiral symmetry of QCD and its 
spontaneous breaking.
The interplay of heavy quark symmetries and chiral symmetry has been 
formulated in terms of effective lagrangians for heavy meson fields
\cite{Donoghue,Wise,Casalbuoni}. 
An effective theory describing the low--energy interactions of heavy 
mesons and heavy baryons with Goldstone bosons $\pi,K,\eta$ was 
discussed in ref.~\cite{Cheng}.
Their results were applied to strong and semileptonic decays using 
spin wave functions of a non--relativistic quark model.
The construction of chiral perturbation theory for heavy hadrons is 
also discussed in ref.~\cite{Cho} where the chiral logarithmic 
corrections to meson and baryon Isgur--Wise functions are given.\\
In ref.\ \cite{EFFR94} (see also \cite{NRZ,BH}) we have presented an 
extension of the NJL quark model, consistent with chiral symmetry for 
the light quarks and the heavy quark symmetries, from which we derived
an effective heavy meson lagrangian and predicted its parameters.
In the present paper we shall
use this model for the description of heavy baryons.

The formation of baryons as bound states of quarks
can be studied in general in 
two complementary pictures. 
One is based upon the limit of infinite number of colors ($N_C$) in QCD,
where baryons arise as solitons of the meson fields \cite{Witten}. 
In fact, heavy baryons have been recently \cite{GWZR95} described as 
a bound state of a chiral 
soliton and a  heavy meson within the extended
(bosonized) NJL model of \cite{EFFR94}.
This so--called bound state approach \cite{CK85} has also been applied 
to a description of heavy baryons within the Skyrme model
\cite{ScheWe95}.
In ref.~\cite{DJM94}, the large 
$N_C$ limit has been analyzed and  an induced algebra with respect to the 
spin--flavor symmetry was derived. 
This model--independent approach was illustrated
in the non--relativistic quark model and the static $SU(2)$ chiral 
soliton model.
Furthermore, in ref. \cite{LR94} a new formalism for treating baryons in the 
$1/N_C$ expansion, based on an analysis of quark--level diagrams, 
has been given.

On the other hand, for finite number of colors ($N_C = 3$), two of 
the three quarks in a baryon are conveniently considered as diquarks.
Then baryons can be described as bound states of quarks and 
diquarks, resulting in Faddeev type of equations
\cite{Ca89,Re90,Eb91}.
Here diquarks play a similar role as the constituent quarks,
 i.e.~they serve as effective (colored) degrees of freedom for the low 
energy dynamics inside the hadron.

In ref.~\cite{Re90} the phenomenologically successful
Nambu--Jona-Lasinio model 
has been converted into an effective hadron theory,
where meson and diquark fields are built from 
correlated $\bar q q$ and $q q$ pairs, respectively,
and baryon fields are constructed as bound quark--diquark states.
This approach, which is fully Lorentz covariant, also allows
to study the connection between the soliton and
the quark--diquark picture of baryons as well as the synthesis of
both pictures \cite{R1}.

In this paper we will apply the above mentioned NJL model with 
heavy quarks to a quark--diquark description of heavy baryons
in lowest order of the $(1/m_Q)$ expansion.
For this aim we will generalize the 
hadronization approach of ref.~\cite{Re90} to the 
lowest--lying baryon states containing a single (infinitely) 
heavy quark, given by the following flavor multiplets \cite{F92,MRR,KKP}:
\begin{description}
\item[i)] 
  Three spin--1/2 baryons where the light quark spins are 
  coupled to zero and which transform as $\bar{3}_F$ 
  with respect to the light flavor group $SU(3)_F$, 
  represented by the antisymmetric matrix:
\be
\label{baryont}
\Pp \ 
\frac{i}{\sqrt{2}}
\left(
\begin{array}{ccc}
0     & \Lambda_Q & \Xi_{Q,\,I_3 = 1/2} \\
- \Lambda_Q & 0         & \Xi_{Q,\,I_3 = -1/2} \\
- \Xi_{Q,\,I_3 = 1/2} & - \Xi_{Q,\,I_3 = -1/2} & 0
\end{array}
\right) \ .
\ee
Here $v^\mu$ is the (conserved) velocity of the
heavy baryon, and $(1 + \slash v)/2$ is a projector
on the large components of the baryon spinor.  
The two spin orientations of the spinors form
  the (in this case trivial) spin-symmetry partners. 
\item[ii)] 
  Six spin multiplets containing a spin--1/2 baryon
  $B_v$ and a spin--3/2 Rarita--Schwinger field
  ${B^*_v}^{\mu}$ transforming as $6_F$ under 
  $SU(3)_F$: 
\ba
&& \frac{1}{\sqrt{3}} \, \gamma_5 \, (\gamma^\mu - v^\mu)
B_v + {B^*_v}^{\mu}   \ , \quad
v_\mu {B_v^*}^\mu = 0
\label{transv}
\\
&& {B_v^{(*)}} =  \Pp \ \frac{1}{\sqrt{2}}\left(
\begin{array}{ccc}
\sqrt{2} \Sigma^{(*)}_{Q,\,I_3 = 1}
& \Sigma^{(*)}_{Q,\,I_3 = 0}
& \Xi^{'(*)}_{Q,\,I_3 = 1/2}\\
\Sigma^{(*)}_{Q,\,I_3 = 0}
& \sqrt{2} \Sigma^{(*)}_{Q,\,I_3 = -1}
& \Xi^{'(*)}_{Q,\,I_3 = -1/2} \\
\Xi^{'(*)}_{Q,\,I_3 = 1/2}
& \Xi^{'(*)}_{Q,\,I_3 = -1/2}
& \sqrt{2} \Omega_Q^{(*)}
\end{array}
\right)
\label{baryons} \ .
\ea
\end{description}
Heavy flavor symmetry leads then to a degeneracy of the residual 
masses $\Delta M = M - m_Q$ for different heavy flavors.\\
The paper is organized as follows:\\
In section \ref{NJL} we define our model which follows  from the
extended Nambu--Jona-Lasinio model of ref.~\cite{EFFR94}
by keeping those
parts relevant for the formation of heavy baryons.
Furthermore, using the above quoted hadronization approach  
our model is transformed into an effective theory for heavy baryons.
The resulting Faddeev equation is analyzed in section \ref{fadeq}, where
we also introduce some approximations which facilitates its numerical
solution.
Numerical results for the heavy baryon masses are presented
in section \ref{results}. Some technical details are relegated to the
appendices.
%
%
%
\section{NJL model with heavy quarks}
\label{NJL}
\subsection{Quark lagrangian and generating functional}
\label{lagrangian}
Let us consider a QCD--motivated NJL model with color--octet
 current--current interaction: 
\be
\L{NJL}^{int} \ =  \ - 3 \ G \ j^A_\mu  \ {j^A}^\mu  \label{Lnjl} \ ,
\ee
where 
$j^A_\mu = \bar \Psi \gamma_{\mu} \frac{\lambda_c^A}{2} \Psi$ , and
$\Psi^T = (q, \ Q)$ contains both the light $q = u,d,s$
and the heavy quarks $Q = c,\, b$, and $G$ is an effective 
coupling constant of dimension mass$^{-2}$.
This interaction can be Fierz--rearranged so that it acts only
in the attractive color singlet quark--antiquark and color anti-triplet
quark--quark channels \cite{Re90}.
These channels can be considered as the physical ones in the following
sense: 
Mesons are built up from color singlet quark--antiquark pairs, while
in a baryon two of the three quarks form a color anti-triplet state.

The interaction is chirally invariant. For the light quark flavors,
the quark--antiquark interaction leads to a spontaneous breaking
of chiral symmetry,
which is accompanied by a dynamical generation of a constituent
quark mass \cite{EbRe86}. Further contributions from the mesonic 
sector will not be taken into account for the calculation of heavy baryons.
Our effective (constituent) quark lagrangian is then defined by:
\be
\label{lag0}
{\cal L} \  =  \ 
\bar q \,  (i \slash{\partial} - m) \,  q
\ + \ 
\bar Q_v \, ( i v \!\cdot\! \partial) \,  Q_v 
 \   + \ {\cal L}^{int} \ ,
\ee
where ${\cal L}^{int} $  represents the diquark correlations
arising from the Fierz transformation of ${\cal L}_{NJL}^{int}$
(\ref{Lnjl}). Since we are interested in baryons with a single
heavy quark we will keep only the light--light and light--heavy
diquark correlations, i.e.\ we ignore diquark correlations with
two heavy quarks.
Then the diquark interaction\footnote{
Heavy quark symmetry and chiral symmetry allow for different 
coupling constants $G_1$, $G_3$ for the light and heavy diquark sector.
A factor 2 is absorbed into the vertices $\Gamma_v$.}
reads:
\bes
{\cal L}^{int} \ = \ 
  2 \, G_1 \,  (\bar q^c  \, \Gamma^{\al} \, q ) \, 
 (\bar q  \, \Gamma^{\al} \,  q^c) 
 \ + \ 
 G_3 \, (\bar q^c \, \Gamma_v^{\alo} \, Q_v)  
 \, (\bar Q_v \, \Gamma_v^{\alo} \, q^c) 
\ . 
\ees
Here, $q^c$ and $\bar q^c$ are the charge conjugated spinors,
given in the  Dirac representation by
\bes
q^c = C \, \bar q^T, \quad \bar q^c = q^T C \ , \quad 
C = i \gamma_2 \gamma_0  = - C^{\da} \ .
\ees
Furthermore, the vertices $\Gamma$ of the diquark currents are
defined by:
\ba
\Gamma^{\al} & := & \left\{ t^a \quad \frac{i\eps^A}{\sqrt{2}} 
 \quad O^{\as} \right\} 
\quad \mbox{(light diquarks)} , \no \\
\Gamma_v^{\alo} & := & \left\{ \quad \frac{i\eps^A}{\sqrt{2}} 
 \quad O_v^{\as} \right\} 
\quad \mbox{(heavy diquarks)} , \no \\
 O^{\as} & := & \left\{ 1, \  i \gamma_5, \  \frac{i}{\sqrt{2}}
 \gamma_{\mu}, \ \frac{i}{\sqrt{2}} \gamma_{\mu} \gamma_5 \right \}
\no \ , \\
 O_v^{\as} & := & \left\{ 1, \  i \gamma_5, \  
 i (\gamma_{\mu} - \slash{v} v_\mu), \ 
 i (\gamma_{\mu} - \slash{v}v_\mu) \gamma_5  \right \} 
\label{gammaH} \ , 
\ea
where $i \epsilon^A$ denotes the color Clebsch-Gordan coefficients
(given by the Levi-Civita tensor $(\epsilon^A)_{ij} = \epsilon_{Aij}$)
coupling the product representation
$3\times 3$ to $3^*$.
Moreover, $t^a   =  \frac{\lambda^a}{2}  , \ a = 1,\ldots,8$  are the generators
of the light flavors $SU(3)_F$ group, and 
$t^0 = \frac{1}{2}\,\sqrt{\frac{2}{3}}\,1_F$. The symbol  $O^\as$ denotes the
Dirac matrices, where in the heavy sector the diquark degrees of freedom have
been rearranged to make vector and axial vector diquarks transversal
with respect to the heavy quark velocity $v^\mu$.\\
Due to the Pauli principle , the light--light diquarks
in the (symmetric) $6_F$ flavor representation exist only
as Lorentz axial--vectors, i.e.~with $O^\as = i \gamma_\mu / \sqrt{2}$,
whereas the diquarks in the (antisymmetric) $\bar 3_F$ representation
occur with $O^\as = \{ 1, \  i \gamma_5,  \ i \gamma_\mu \gamma_5 \}$.
For the $\Gamma_v$ vertices of the heavy--light
 diquarks\footnote{In the following we will refer
to the  diquark fields simply as light and heavy diquarks,
respectively.}
there is no restriction since heavy and light quarks have to be treated
as different particles.
\subsection{Hadronization}
Our aim is to convert the quark theory defined by (\ref{lag0}) into an 
effective theory of heavy baryons.
For this purpose we follow the hadronization approach of \cite{Re90},
where baryon fields were built from diquark and quark fields.

Baryons with a heavy quark can be constructed from either a light
diquark and a heavy quark or a heavy diquark and a light quark.
Therefore we introduce
both types of diquark fields into the generating functional
\bes
Z  \ = \  \int D \bar q \ D q \ D \bar Q_v \ D Q_v \ 
      \exp\left[ i \int  {\cal L} \right] 
\ees
with the help of the following 
identities\footnote{We have defined the
conjugated fields in such a way that 
$
 \left( \Delta \, \Gamma \, \gamma_0 \right )^\da :=
 \Delta^\da \, \Gamma \, \gamma_0
$. 
Since some of the $O^{\as}$ are anti--hermitian, 
the respective diquark fields behave likewise, leading
to an unusual phase convention. \label{foot1}  As usual
irrelevant normalization factors in front of functional integrals
are omitted. The numerical factors $1/2$, $4G_1$, $2G_3$ 
in eqs.~\gl{einsa}-\gl{einsd} have
been introduced for later notational convenience.
In order to keep notations transparent,
 we sometimes write $\int$ instead of $\int d^4x$ etc.}:
\ba 
1  &=&  \int \, \prod_\al \, D \kappa^\al \, D {\kappa^\da}^ \al 
\,  \delta ({\kappa^\da}^\al  - \bar q  \, \Gamma^\al \,  q^c)
\,  \delta ({\kappa}^\al  - \bar q^c  \, \Gamma^\al \,  q)
 \no \\
& = &  
\int \prod_{\al} \ 
D \kappa^{\al}  \, D {\kappa^{\da}}^{\al} \,
D \Delta^{\al}  \, D {\Delta^{\da}}^{\al} \, 
 \exp \,
 \frac{i}{2} \int \left[ \Delta^{\al} \,
( {\kappa^{\da}}^\al - \bar q \, \Gamma^{\al} \,q^c )  
 \, + \,
  h.c. \right] \ ,
\label{einsa} 
\\
1  &=&  \int \, \prod_\alo \, D \kappa_v^\alo \, D {\kappa_v^\da}^ \alo 
\,  \delta ({\kappa_v^\da}^\alo  - \bar Q_v  \, \Gamma_v^\alo \,  q^c)
\,  \delta ({\kappa_v}^\alo  - \bar q^c  \, \Gamma_v^\alo \,  Q_v)
 \no \\
 &=&  
\int \prod_{\alo} \, 
D \kappa_v^{\alo}  \, D {\kappa_v^{\da}}^\alo \,
D \Delta_v^{\alo}  \, D {\Delta_v^{\da}}^\alo \, 
 \exp \,
 \frac{i}{2} \int \left[ \Delta_v^{\alo} \,
( {\kappa_v^{\da}}^\alo - \bar Q_v \, \Gamma^{\alo}_v \, q^c )  \, 
   \, + \,
  h.c. \right] 
\label{einsb} \ .
\ea 
Note that the heavy diquark field 
$\Delta_v  = \left (\Delta_v \right)_j$ (and $(\kappa_v^\da)_j$) 
with $j=u,d,s$ transforms as  a triplet under $SU(3)_F$.
Analogously, we also introduce two different baryon amplitudes for the 
heavy baryons:
A baryon field amplitude ${\cal X}_v$ built up from a light diquark and a heavy
quark (${\cal X}_v \sim (\bar q^c q) \, Q_v$) and an amplitude 
${\cal Y}_v$ constructed from a heavy diquark and a light quark (${\cal Y}_v
 \sim (\bar q^c Q_v) \, q )$.\\
These fields are introduced again by inserting the following identity 
into the generating functional:
\ba
1 & = & \int \, \prod_{\al} \, D \bar X_v^{\al} \, D  X_v^{\al} 
        \, D \bar {\cal X}_v^{\al}\, D {\cal X}_v^{\al} \cdot
\no \\
& &  \cdot \exp \, i \, \int \left\{ 
 \left[ (\bar {\cal X}_v^{\al} (x,y)  - 
 4 \, G_1 \, \bar Q_v(x) \, 
 [ \bar q \, (y) \, \Gamma^{\al} \,  q^c(y) ] \, \right ] 
 \, X_v^{\al}(x,y)  + h.c. \right\} 
\label{einsc} \ ,
\\
1 & = & \int \,  \prod_{\alo, \, ij}\,  
D {\bar Y_v}{}^{\alo}_{ij} \, D {Y_v}{}^{\alo}_{ij} 
 \, D {\bar {\cal Y}_v}{}^{\alo}_{ij} \, D {{\cal Y}_v}{}^{\alo}_{ij} 
\cdot \no \\
& & \cdot \exp \, i \, \int \left\{ \left[
 {\bar{\cal Y}_v}{}^{\alo}_{ij} (x,y) - 2\, G_3 \, \bar q_i(x) \,  
  [ \bar Q_v (y) \, \Gamma_v^{\alo} \,  q_j^c(y) ]\, \right] 
  \,  {Y_v}^{\alo}_{ij}(x,y) + h.c. \right\} \ . 
\label{einsd}
\ea 
Note also that $\slash{v} {\cal X}_v = {\cal X}_v$ and
$v_\mu {\cal Y}_v^\mu = 0$.
For later convenience, we have introduced the baryon fields
in two different flavor bases: The flavor index $a$ of ${\cal X}^\al_v$
 (contained in $\al = \{a, A,\as \}$), 
$a = 0,\ldots,8$, refers to the generators of the $SU(3)_F$ group and
the unit matrix,
whereas the flavor indices ${i,j}$ of
 ${\cal Y}_v{}^{\alo}_{ij}$  ($\alo = \{A, \as\}$),
$i,j = 1..3$, denote the light flavor of the quark and the heavy diquark,
respectively.

Owing to the $\delta$-functionals in (\ref{einsa})--(\ref{einsd}),
we can replace the quark bilinears in  the interaction (\ref{lag0})
 by the fields $\kappa,\, \kappa_v$ and 
$\kappa^\dagger,\, \kappa_v^\dagger$,
resulting in a lagrangian which is bilinear in the quark fields.\\
After integrating out the quark fields as well as the
 auxiliary fields $\kappa$, we 
 obtain an effective theory in terms of the diquark fields
 $\Delta, \Delta_v$ and the baryon fields $X_v, {\cal X}_v, Y_v , {\cal Y}_v$
(see (\ref{seff})).
This effective baryon theory is highly non--local and we will
work it out in the low--energy regime.
Since we are interested in the description of individual baryons
we expand the corresponding effective action
up to second order in the baryon source fields\footnote{
Higher powers of baryon fields in (\ref{seffbar})
 would lead to the generation of
baryon interactions in the effective lagrangian
which are outside the scope of this paper.} $X_v^\al, Y_v^{\al_0}$.
This yields (see appendix A for more details):
\ba
Z & = & \int \ D \Delta \ D \Delta^\da  \ D \Delta_v \ D \Delta_v^\da \ D X_v 
\ D \bar X_v \cdots
 \ \exp \ \left ( i S_{\rm eff} \right ) \no \ , \\
S_{\rm eff}  &=&
 - \int \,\frac{1}{8\,G_1} \, \Delta^{\al} \, {\Delta^{\da}}^\al
 - \int \,\frac{1}{4\,G_3} \,{\Delta_v^\alo} \, {\Delta_v^{\da}}^\alo 
 + \int \left( \bar {\cal X}_v^{\al} X_v^\al  
 + {\bar {\cal Y}_v}{}_{ij}^\alo {Y_v}_{ij}^\alo + h.c.\right)
\no \\
&  & 
- i  \, {\rm Tr} \, \log \,  g_v^{-1}  - i  \, {\rm Tr} \, \log \,  g^{-1}  
- \frac{i}{2}  \, {\rm Tr} \, \log 
    ( 1  -  g \,\Delta \,\tilde g \,  \Delta^\da)
\no \\
&&
 -  \int \,( \bar X^\bt_v ,  \, \bar {Y_v}^\bto_{ij} )  \,
\left( \begin{array}{cc}
(G_{B\ XX})^{\bt\bt'} & (G_{B\ XY})^{\bt\bto'}_{\ i'j'} \\
(G_{B\ YX})^{\bto\bt'}_{ij} & (G_{B\ YY})^{\bto\bto'}_{ij\,i'j'}
\end{array} \right) \,
   \left ( \begin{array}{c} X^{\bt'}_v \\ {Y_v}^{\bto'}_{i'j'}
 \end{array}
 \right ) \ .  \label{seffbar} 
\ea
For this purpose we have defined the inverse of the light and heavy quark
propagators as:
\ba
g^{-1}(x,y) & = & 
( i \slash{\partial}_x - m) \, \delta(x-y) 
+ \frac{1}{4} \, \left[ C \,  \Delta_v^\da (y)
\, g_v(y,x) \, \Delta_v (x)\, C \right ]^T
\ , \quad  \tilde g  \ = \  C^\da g^T C 
 \ , \\
g_v^{-1}(x,y) &=& \Pp \, (i v \!\cdot\! \partial_x) \,
        \delta(x-y) \ ,   \qquad
        \tilde g_v \ = \  C^\da g_v^T C \ , \label{gv}
\ea
with the abbreviations
$
 \ \Delta := \Delta^\al \, \Gamma^\al ,  
 \ \Delta^\da := {\Delta^\da}^\al
 \Gamma^\al  
$, 
$
 \ {\Delta_v}  := {\Delta_v}^\alo \, \Gamma_v^\alo ,  
 \ {\Delta_v^\da} := {\Delta_v^\da}^\alo \, \Gamma^\alo  
$.
The symbol Tr denotes the trace over coordinate space and flavor, 
color and Dirac indices. 
The entries for the matrix $G_B$ 
are of $0^{th}$ order in the baryonic fields
$X_v$ and $Y_v$, but still contain all orders
 of quark--diquark contributions, see (\ref{gyy}), (\ref{g22}).
After the Gaussian integration over the auxiliary baryonic fields
$X_v$ and $Y_v$, the matrix $G_B$ can be identified as the propagator
for the baryon fields ${\cal X}_v$ and ${\cal Y}_v$.

\subsection{Derivation of the baryon propagator}
\label{barprop}
To obtain the effective theory of the heavy baryon fields,
we have to integrate out the diquark fields $\Delta$.
This can only be done  in an approximate fashion
due to the presence of the quark loops coupling to the diquark fields
in the effective action (\ref{seffbar}).
First we derive the free diquark propagators $D, D_v$  in leading order 
of the loop
expansion, i.e.~we expand the term Tr log $(1 - g \ \Delta \ \tilde g \
  \Delta^\da)$
in (\ref{seffbar}) up to second order in the diquark fields. 
This yields:
\ba
 {D^{\bt \bt'}}^{-1}(x,y)  & = &   
- \frac{1}{8 G_1} \ \delta^{\bt \bt'} \ \delta(x-y)
\ + \ \frac{i}{2} \ \tr{\Gamma^{\bt} \ g_0(x,y) \ \Gamma^{\bt'}
\ \tilde g_0(y,x)} \label{ldiq} \\
{(D_v)^{\bto \bto'}_{jj'}}^{-1}(x,y)  & = & 
 - \frac{1}{4 G_3} \ \delta^{\bto \bto'} \
\delta_{jj'} \ 
\delta(x-y) \ + \ 
\frac{i}{4} \ \tr{ \Gamma_v^{\bto} \ 
 g_v(x,y) \ \Gamma_v^{\bto'} \  (\tilde g_0)_{jj'}(y,x)}
\label{hdiq}
\ea
for light and heavy diquarks, respectively, with
$(g_0)_{ii'}(x,y) := \left (i\slash\partial_x - m_i \right)^{-1} 
        \delta(x-y) \, \delta_{ii'}$ denoting the free light
quark propagator.
The diquark polarization tensors (the terms in the square brackets) are
 calculated  in appendix \ref{B} and graphically depicted in
 Fig.~\ref{fig-diq}. 
Note that the heavy diquark propagator is a diagonal
flavor matrix,  
$\left\{D_v \right\}_{jj'} = {\rm diag} (D_v^u, D_v^d, D_v^s)$.
\begin{fmffile}{pic}
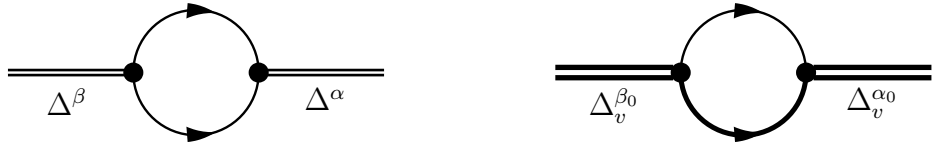
\begin{figure}[htb]
\begin{center}
\unitlength1.0cm

\begin{fmfgraph*}(5,3)
\fmfpen{thick}

\fmfleft{in}
\fmfright{out}

\fmf{dbl_plain,width=thin,label=\noexpand $\Delta^\beta$}{in,v1}
\fmf{dbl_plain,width=thin,label=\noexpand $\Delta^\alpha$}{v2,out}

\fmf{fermion,tension=0.5,width=thin,right}{v1,v2}
\fmf{fermion,tension=0.5,width=thin,left}{v1,v2}

\fmfv{de.sh=circle,de.si=0.04w,de.fi=1}{v1}
\fmfv{de.sh=circle,de.si=0.04w,de.fi=1}{v2}
\end{fmfgraph*}
\hspace{2cm}
\begin{fmfgraph*}(5,3)
\fmfpen{thick}
\fmfleft{in}
\fmfright{out}

\fmf{dbl_plain,width=thick,label=\noexpand $\Delta_v^{\beta_0}$}{in,v1}
\fmf{dbl_plain,width=thick,label=\noexpand $\Delta_v^{\alpha_0}$}{v2,out}

\fmf{fermion,tension=0.5,width=thick,right}{v1,v2}
\fmf{fermion,tension=0.5,width=thin,left}{v1,v2}

\fmfv{de.sh=circle,de.si=0.04w,de.fi=1}{v1}
\fmfv{de.sh=circle,de.si=0.04w,de.fi=1}{v2}

\end{fmfgraph*}
\end{center}
\caption{Diquark polarization tensors for light and heavy
diquarks, generated by the loop expansion
of the quark determinant.}
\label{fig-diq}
\end{figure}

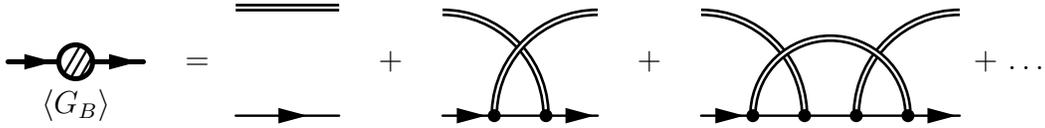
\begin{figure}[t]
\begin{center}
\unitlength1.0cm

\ba
\nonumber
\parbox{20mm}{

\begin{fmfgraph*}(1.8,2)
\fmfpen{thick}
\fmfleft{d1,i,d2}
\fmfright{d3,f,d4}
\fmf{fermion}{i,a}
\fmf{fermion}{a,f}
\fmfblob{.25w}{a}
\fmf{phantom,l.side=left,label=$\noexpand \langle G_B \noexpand \rangle$}{d1,d3}
\fmf{phantom}{d2,d4}
\end{fmfgraph*}
}
&=&
\parbox{110mm}{

\begin{fmfgraph*}(11,2.3)
\fmfpen{thin}

\fmfforce{(0,3/16 h)}{qi}
\fmfforce{(0,13/16 h)}{Di}
\fmfforce{(w,3/16 h)}{qf}
\fmfforce{(w,13/16 h)}{Df}
\fmfforce{(0,1/2 h)}{mi}
\fmfforce{(w,1/2 h)}{mf}

\fmf{phantom}{qi,q2,q3,q4,q5,q6,q7,q8,q9,q10,q11,q12,q13,q14,q15,q16,qf}
\fmf{phantom}{Di,D2,D3,D4,D5,D6,D7,D8,D9,D10,D11,D12,D13,D14,D15,D16,Df}
\fmf{phantom}{mi,m2,m3,m4,m5,m6,m7,m8,m9,m10,m11,m12,m13,m14,m15,m16,mf}

\fmffreeze

\fmf{fermion}{qi,q3}
\fmf{fermion}{q5,q6}
\fmf{plain}{q6,q7}
\fmf{fermion}{q7,q8}
\fmf{fermion}{q10,q11}
\fmf{plain}{q11,q12,q13,q14}
\fmf{fermion}{q14,q15}

\fmf{dbl_plain}{Di,D3}
\fmf{dbl_plain,left,tension=1.5}{q11,q14}

\fmfdot{q6,q7,q11,q12,q13,q14}
\fmfv{label=$+$,l.a=0,l.d=0}{m4}
\fmfv{label=$+$,l.a=0,l.d=0}{m9}
\fmfv{label=$+$ \noexpand \ldots,l.a=0,l.d=0}{m16}
\fmf{phantom,right,tension=2,tag=1}{q7,q3}
\fmf{phantom,left,tension=2,tag=2}{q6,q10}
\fmf{phantom,right,tension=2,tag=3}{q12,q8}
\fmf{phantom,left,tension=2,tag=4}{q13,qf}

\fmfposition

\fmfipath{p[]}
\fmfiset{p1}{vpath1(__q7,__q3)}
\fmfiset{p2}{vpath2(__q6,__q10)}
\fmfiset{p3}{vpath3(__q12,__q8)}
\fmfiset{p4}{vpath4(__q13,__qf)}

\fmfi{dbl_plain}{
        subpath (0,length(p1)/2) of p1}

\fmfi{dbl_plain}{
        subpath (0,length(p2)/2) of p2}

\fmfi{dbl_plain}{
        subpath (0,length(p3)/2) of p3}

\fmfi{dbl_plain}{
        subpath (0,length(p4)/2) of p4}

\end{fmfgraph*}}
\ea
\end{center}
\caption{Diagrammatic expansion of the baryon propagator.}
\label{fig-barprop}
\end{figure}

\begin{figure}[thb]
\begin{eqnarray*}
\mbox{a)} \
\parbox{2.4cm}{
\begin{center}
\unitlength1.2cm

\begin{fmfgraph}(1,3)

\fmfbottom{qin,Din}
\fmftop{qout,Dout}

\fmf{fermion}{qin,qout}
\fmf{dbl_plain}{Din,Dout}
\end{fmfgraph}
\end{center}}
\quad
\mbox{b)} \
\parbox{2.4cm}{
\begin{center}
\unitlength1.2cm

\begin{fmfgraph}(1.5,3)

\fmfbottom{qin,Din}
\fmftop{qout,Dout}

\fmf{fermion}{qin,qout}
\fmf{dbl_plain}{Din,v1}
\fmf{fermion,left,tension=0.3}{v1,v2}
\fmf{fermion,right,tension=0.3}{v1,v2}
\fmf{dbl_plain}{v2,Dout}
\fmfdot{v1,v2}
\end{fmfgraph}
\end{center}}
\quad
&\leftrightarrow&
\
\mbox{c)} 
\parbox{2.4cm}{
\begin{center}
\unitlength1.2cm

\begin{fmfgraph}(1.5,3)

\fmfbottom{qin,Din}
\fmftop{qout,Dout}

\fmf{phantom}{qin,qout}
\fmf{phantom}{Din,v1}
\fmf{phantom,tension=0.5}{v1,v2}
\fmf{phantom}{v2,Dout}

\fmffreeze

\fmf{fermion}{qin,v2}
\fmf{dbl_plain}{Din,v1}
\fmf{fermion}{v1,v2}
\fmf{fermion}{v1,qout}
\fmf{dbl_plain}{v2,Dout}
\fmfdot{v1,v2}

\end{fmfgraph}
\end{center}}
\quad
=
\
\mbox{d)} \
\parbox{2.4cm}{
\begin{center}
\unitlength1.2cm

\begin{fmfgraph}(1.5,3)

\fmfbottom{qin,Din}
\fmftop{qout,Dout}

\fmf{phantom}{qin,v1}
\fmf{phantom,tension=0.5}{v1,v2}
\fmf{phantom}{v2,qout}
\fmf{phantom}{Din,Dout}

\fmffreeze

\fmf{fermion}{qin,v1}
\fmf{fermion}{v2,v1}
\fmf{fermion}{v2,qout}
\fmf{dbl_plain}{v1,Dout}
\fmf{dbl_plain}{Din,v2}
\fmfdot{v1,v2}

\end{fmfgraph}
\end{center}}
\end{eqnarray*}
\caption{a) The unperturbed baryon propagator describing
the independent propagation of a quark and a diquark.
By construction of the diquark propagator, this diagram
includes already the diagram shown in b). \ c)
Exchange diagram to b). \ d) 
The exchange diagram c) redrawn to show the equivalence to
 Fig.~\ref{fig-barprop}. }
\label{fig-exc}
\end{figure}
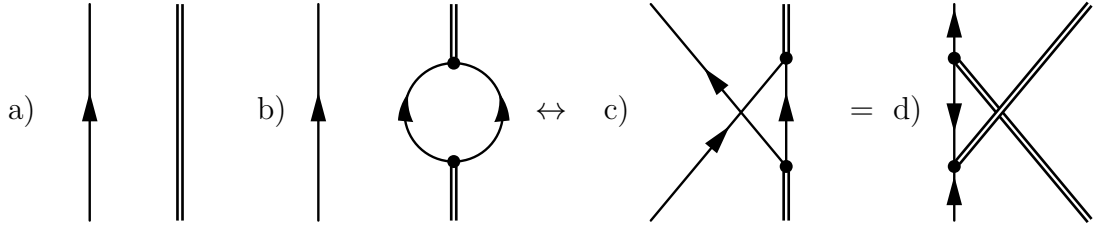

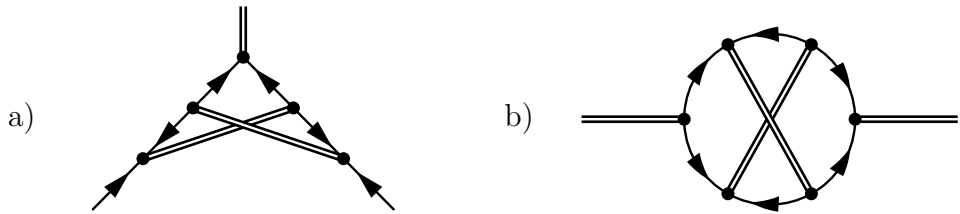
\begin{figure}[thb]
\begin{center}
\mbox{a)} \
\parbox{5cm}{
\begin{center}
\unitlength1.0cm

\begin{fmfgraph}(4,3)

\fmfbottom{q1,q2}
\fmftop{Dout}

\fmf{fermion}{q1,v1}
\fmf{fermion}{q2,v2}
\fmf{fermion}{v3,v1}
\fmf{fermion}{v4,v2}
\fmf{fermion}{v4,v}
\fmf{fermion}{v3,v}
\fmfdot{v,v1,v2,v3,v4}
\fmf{dbl_plain,tension=2}{v,Dout}
\fmffreeze
\fmf{dbl_plain}{v1,v4}
\fmf{dbl_plain}{v2,v3}
\end{fmfgraph}
\end{center}}
\hskip2em
\mbox{b)} 
\parbox{6cm}{
\begin{center}
\unitlength1cm
\begin{fmfgraph}(5,3)
\fmfleft{in}
\fmfright{out}

\fmf{dbl_plain}{in,v1}
\fmf{dbl_plain}{v2,out}

\fmf{phantom,tension=0.3,width=thin,right,tag=1}{v1,v2}
\fmf{phantom,tension=0.3,width=thin,left,tag=2}{v1,v2}
\fmfdot{v1,v2}
\fmffreeze
\fmfposition

\fmfipath{p[]}
\fmfiset{p1}{vpath1(__v1,__v2)}
\fmfiset{p2}{vpath2(__v1,__v2)}

\fmfi{fermion}{subpath (0,length(p1)/3) of p1}
\fmfi{fermion}{subpath (2 length(p1)/3,length(p1)/3) of p1}
\fmfi{fermion}{subpath (2 length(p1)/3,length(p1)) of p1}
\fmfi{fermion}{subpath (0,length(p2)/3) of p2}
\fmfi{fermion}{subpath (2 length(p2)/3,length(p2)/3) of p2}
\fmfi{fermion}{subpath (2 length(p2)/3,length(p2)) of p2}

\fmfforce{point length(p1)/3 of p1 }{a1}
\fmfforce{point 2 length(p1)/3 of p1 }{a2}
\fmfforce{point length(p2)/3 of p2 }{a3}
\fmfforce{point 2 length(p2)/3 of p2 }{a4}

\fmf{dbl_plain}{a1,a4} \fmf{dbl_plain}{a2,a3}
\fmfdot{a1,a2,a3,a4}

\end{fmfgraph}
\end{center}}
\end{center}
\caption{a) A typical higher-order correction to the quark-diquark vertex.
        b) Corresponding higher-order correction to the diquark
        propagator, which is discarded in this paper.}
\label{fig-higher}
\end{figure}

Now  the functional 
integration over $\Delta$ and $\Delta^\da$ will be performed
 in leading order of the cluster expansion \cite{VanKampen}:
\be
\langle \exp F \rangle_\Delta \ = \ \exp \left( \sum_{n=1}^\infty 
                \frac{1}{n!}\, \langle\langle F^n \rangle\rangle \right)
= \exp\left\{ \langle F \rangle_\Delta + O(\langle F^2
\rangle_\Delta - \langle F \rangle_\Delta^2 ) \right\} \ ,
\ee
where the functional average over the diquark fields is defined by:
\bes
\langle \left[\dots\right] \rangle_\Delta   
 \ =\    \frac{ \int D \Delta^{\da} D \Delta 
 D \Delta_v^{\da} D \Delta_v 
 \left[\dots\right] \exp \  i  \int
 \left(
  {\Delta^\da}^{\al} (D^{-1})^{\al \bt} \Delta^{\bt} 
 +
  {\Delta_v^\da}^{\alo} (D_{v}^{-1})^{\alo \bto} \Delta_v^{\bto} 
 \right)}
 {\int D \Delta^{\da} D \Delta D \Delta_v^{\da} D \Delta_v  \  
  \exp \  i  \int
 \left(
 {\Delta^\da}^{\al} (D^{-1})^{\al \bt} \Delta^{\bt} 
+
 {\Delta_v^{\da}}^\alo (D_{v}^{-1})^{ \alo \bto} \Delta_v^{\bto} 
\right) } \ .
\ees
Obviously, the  diquark propagators (\ref{ldiq}) and 
(\ref{hdiq}) are then given by:
\ba
i D^{\bt \bt'}(x,y)  & = & 
   \langle \Delta^{\bt}(x) {\Delta^\da}^{\bt'}(y) \rangle \ ,  \\
i (D_v)_{jj'}^{\bto \bto'}(x,y)  & = & 
   \langle {\Delta_v}^{\bto}_j(x) {\Delta_v^\da}^{\bto'}_{j'}(y) \rangle
 \ . 
\ea
Now we are left  with the task to evaluate the functional average
 $\langle G_B \rangle$ over the diquark fields.
With the help of Wick's theorem this 
functional average can be reformulated into the sum over all possible
contractions of the diquark fields in $G_B$ into free 
diquark propagators $\langle \Delta \Delta^\da \rangle$.
Hereby, diagrams with an arbitrary number of intermediate diquark
propagators arise. This reflects the fact that even in the
constiuent quark model the baryon is in principle a 
many-particle system, as soon as the Dirac sea is included.
In the following, we will only keep the simply nested exchange
diagrams shown in Fig.~\ref{fig-barprop}.
Here the first diagram describes the independent propagation of
a quark and a diquark. A consistent treatment requires to include
also the corresponding exchange diagram shown in Fig.~\ref{fig-exc}.
Notice that there arise also self-energy corrections for the
quark propagator due to inserted diquark propagators which can
be absorbed into the constituent quark mass. Moreover, there
emerge higher order contributions to the exchange diagram
in Fig.~\ref{fig-exc}c) due to corrections to the quark-diquark
vertex (see Fig.~\ref{fig-higher}a).
A consistent treatment of the baryon would then require
to include the same correction into the diquark propagator.
This complicated procedure which goes beyond the one-loop
expression for the diquark propagator is clearly outside
the scope of this paper.

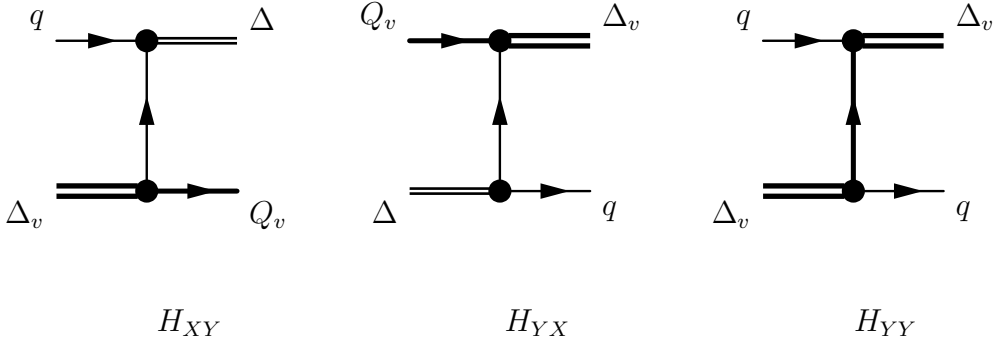
\begin{figure}[thb]
\vskip0.9em
\begin{center}
\unitlength1.0cm
\begin{tabular}{ccc}
\parbox{40mm}{

\begin{fmfgraph*}(3,2)
\fmfpen{thick}

\fmfleft{Di,qi}
\fmfright{qf,Df}

\fmf{fermion,width=thin}{qi,a}
\fmfv{de.sh=circle,de.fi=1,de.si=0.08w}{a}
\fmf{dbl_plain,width=thin}{a,Df}
\fmf{dbl_plain,width=thick}{Di,b}
\fmfv{de.sh=circle,de.fi=1,de.si=0.08w}{b}
\fmf{fermion,width=thick}{b,qf}

\fmffreeze

\fmf{fermion,width=thin}{b,a}

\fmflabel{$\Delta_v$}{Di}
\fmflabel{$\Delta$}{Df}
\fmflabel{$q$}{qi}
\fmflabel{$Q_v$}{qf}

\end{fmfgraph*}
}
\ & \
\parbox{40mm}{

\begin{fmfgraph*}(3,2)
\fmfpen{thick}

\fmfleft{Di,qi}
\fmfright{qf,Df}

\fmf{fermion,width=thick}{qi,a}
\fmfv{de.sh=circle,de.fi=1,de.si=0.08w}{a}
\fmf{dbl_plain,width=thick}{a,Df}
\fmf{dbl_plain,width=thin}{Di,b}
\fmfv{de.sh=circle,de.fi=1,de.si=0.08w}{b}
\fmf{fermion,width=thin}{b,qf}

\fmffreeze

\fmf{fermion,width=thin}{b,a}

\fmflabel{$\Delta$}{Di}
\fmflabel{$\Delta_v$}{Df}
\fmflabel{$Q_v$}{qi}
\fmflabel{$q$}{qf}

\end{fmfgraph*}
}
\ & \
\parbox{40mm}{

\begin{fmfgraph*}(3,2)
\fmfpen{thick}

\fmfleft{Di,qi}
\fmfright{qf,Df}

\fmf{fermion,width=thin}{qi,a}
\fmfv{de.sh=circle,de.fi=1,de.si=0.08w}{a}
\fmf{dbl_plain,width=thick}{a,Df}
\fmf{dbl_plain,width=thick}{Di,b}
\fmfv{de.sh=circle,de.fi=1,de.si=0.08w}{b}
\fmf{fermion,width=thin}{b,qf}

\fmffreeze

\fmf{fermion,width=thick}{b,a}

\fmflabel{$\Delta_v$}{Di}
\fmflabel{$\Delta_v$}{Df}
\fmflabel{$q$}{qi}
\fmflabel{$q$}{qf}

\end{fmfgraph*}
}
\\
&&\\[1cm]
$H_{XY}$ & $H_{YX}$ & $H_{YY}$
\end{tabular}

\end{center}
\caption{Elements of the
interaction matrix $H$.}
\label{fig-hint}
\end{figure}

Let us remark that in the case of light quarks the class of
diagrams shown in Fig.~\ref{fig-barprop} just represents the
minimal subset of diagrams which is required to fulfill the
Pauli-principle in the quark-diquark picture of baryons
\cite{Re90,R1}.
The above series of diagrams defines a kind of ladder
approximation which in the next step will be summed up
to yield a Faddeev type of 3-body amplitude.
Hereby, the presence of both light and heavy diquarks
induces a $ 2\! \times\! 2$ matrix structure. 
Therefore, we introduce the matrix of the free quark--diquark propagation
$G= {\rm diag}\left(G_{XX},\, G_{YY}\right)$, 
\ba
( G_{XX})^{\bt\bt'}(x,x';\, y,y') &  =  & 
 i \ D^{\bt\bt'}(x',y') \ g_v(x,y) 
\ , \no \\
 (G_{YY})^{\bto\bto'}_{ij \ i'j'}(x,x';\, y,y') &  = &  
i (D_v)^{\bto \bto'}_{jj'} (x',y') (g_0)_{ii'} (x,y) 
\ . \label{gfrei}
\ea
Here, $g_v$ and $g_0$
denote the free propagators of heavy and light constituent quarks,
respectively.
Now we define  the interaction matrix $H$,
whose  elements mediate
the transitions  ${\cal X}_v \leftrightarrow {\cal Y}_v$ and ${\cal Y}_v
 \to {\cal Y}_v$. As can be read off from Fig.~\ref{fig-barprop} by
straightening the  diquark lines, these transitions
are mediated by quark--exchange. In Fig.~\ref{fig-hint} the elements
of $H$ are depicted graphically. We have:
\ba
H & = & 
\left ( \begin{array}{cc} 
0 & H_{XY} \\ 
H_{YX} & H_{YY}
\end{array} \right )  
\ , 
\no
\\
(H_{XY})^{\bt \bto'}_{\ \ i'j'}(x,x';\, y,y')  
& = & 
\frac{1}{2} \  
\Gamma_v^{\bto'}  \ (\tilde g_0(x,y) \Gamma^{\bt})_{j'i'} 
\ \delta(x' - y) \ \delta(y' - x) \ ,
\no
\\
(H_{YX})^{\bto \bt'}_{ij}(x,x';\, y,y')  
& = & 
\frac{1}{2} \  
(\Gamma^{\bt'} \tilde g_0(x,y))_{ij} \ 
\Gamma_v^{\bto} \ \delta(x' - y) \ \delta(x - y') \ , 
\no
\\ 
(H_{YY})^{\bto \bto'}_{ij \ i'j'} (x,x';\, y,y') 
& = &    
\frac{1}{4}  
\, \tilde \Gamma_v^{\bto'}  \  \tilde g_v(x,y) \ 
 \tilde \Gamma_v^{\bto} 
 \ \delta(x' - y) \ \delta(x - y') 
 \ \delta_{ji'} \, \delta_{ij'} \ . 
\label{h} 
\ea 

Here, $\tilde \Gamma_v \ = \  C^\da \ \Gamma_v^T \  C$.
Note that $H_{YY}$ formally mediates the exchange of the heavy quark.
With the above definitions of $G$ and $H$,
the baryon propagator defined by the series of diagrams shown in
Fig.~\ref{fig-barprop} becomes:
\be
\langle G_B \rangle  =  
G \  \sum_k ( H \ G )^k   
=   \ G \  ( 1 - H \ G )^{-1} \ . \label{gb}
\ee
It has the standard form of a Faddeev type of 3--particle propagator.
%
\section{The Faddeev equation}
\label{fadeq}
Baryon masses are given by the poles of the baryon propagator
(\ref{gb}), 
which results in a Faddeev type of equation for the baryon
wave functions:
\be 
( 1 -  G \, H )
 \ \left
 ( \begin{array}{c} {\cal X}_v \\ {\cal Y}_v \end{array} \right ) 
  =  0  \ .
\label{faddeev}
\ee
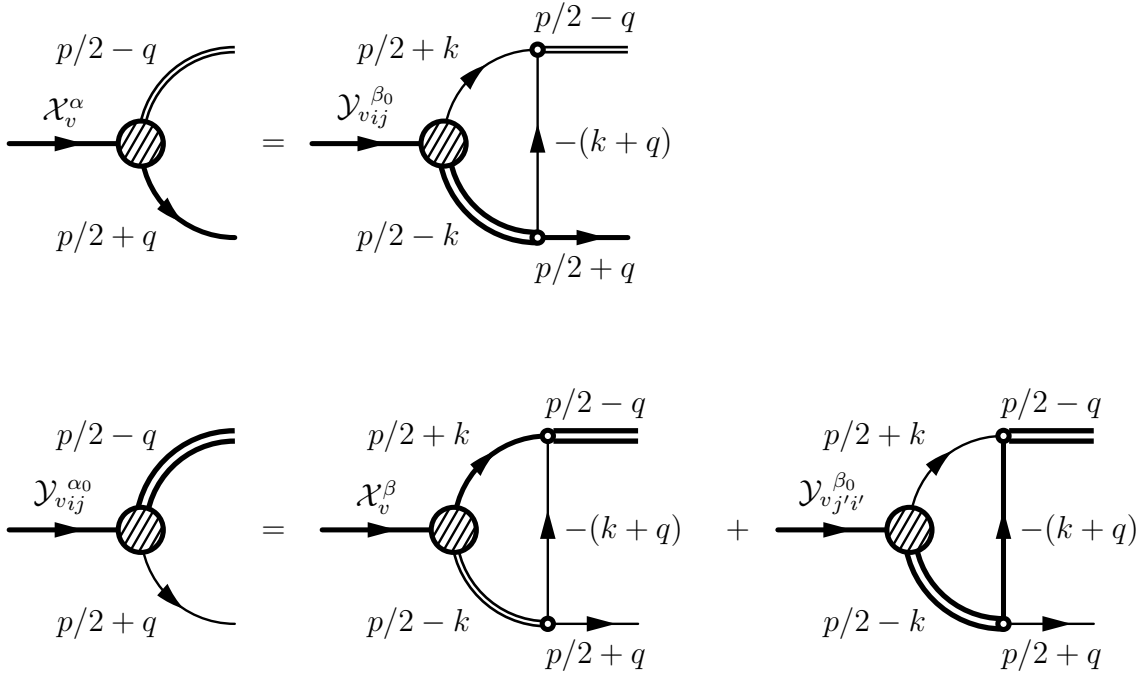
\begin{figure}[tbh]
\unitlength0.6cm
\begin{eqnarray}
\nonumber
\parbox{30mm}{

\begin{fmfgraph*}(10,5)
\fmfpen{thick}

\fmfleft{psi}
\fmfright{psi0}

\fmf{fermion,
        label=${\noexpand \cal X}_v^\alpha$,
        label.side=left,width=thick}{psi,blob}
\fmf{phantom}{blob0,psi0}

\fmf{phantom,right,tension=0.35,tag=1}{blob,blob0}
\fmf{phantom,tension=0.35,tag=2,left}{blob,blob0}

\fmfposition

\fmfipath{p[]}
\fmfiset{p1}{vpath1(__blob,__blob0)}
\fmfiset{p2}{vpath2(__blob,__blob0)}

\fmfi{fermion,
        label=$ p/2 +q$,
        label.side=right,
        width=thick}{
        subpath (0,length(p1)/2) of p1}

\fmfi{dbl_plain,
        label=$p/2 -q$,
        label.side=left,
        width=thin}{
        subpath (0,length(p2)/2) of p2}

\fmffreeze

\fmfforce{point length(p1)/2 of p1 }{q}
\fmfforce{point length(p2)/2 of p2 }{D}

\fmfblob{0.2h}{blob}
\end{fmfgraph*}
}
&=&
\parbox{50mm}
{
\begin{fmfgraph*}(10,5)
\fmfpen{thick}

\fmfleft{psi}
\fmfright{psi0}

\fmf{fermion,
        label=\noexpand ${{\noexpand \cal Y}_v}^{\beta_0}_{ij}$,
        label.side=left,
        width=thick}{psi,blob}
\fmf{phantom}{blob0,psi0}

\fmf{phantom,right,tension=0.35,tag=1}{blob,blob0}
\fmf{phantom,tension=0.35,tag=2,left}{blob,blob0}

\fmfposition

\fmfipath{p[]}
\fmfiset{p1}{vpath1(__blob,__blob0)}
\fmfiset{p2}{vpath2(__blob,__blob0)}

\fmfi{dbl_plain,
        label=$p/2 -k$,
        label.side=right,
        width=thick}{
        subpath (0,length(p1)/2) of p1}

\fmfi{fermion,
        label=$p/2 +k$,
        label.side=left,
        width=thin}{
        subpath (0,length(p2)/2) of p2}

\fmffreeze

\fmfforce{point length(p1)/2 of p1 }{q}
\fmfforce{point length(p2)/2 of p2 }{D}
\fmfforce{point length(p1)/2 of p1 }{q1}
\fmfforce{point length(p2)/2 of p2 }{D1}

\fmffreeze

\fmfshift{(0.2w,0)}{q}
\fmfshift{(0.2w,0)}{D}

\fmffreeze

\fmf{fermion,label=
$-(k+q)$,
     label.side=right,width=thin}{q1,D1}

\fmf{dbl_plain,label=$p/2 - q$, label.side=left,width=thin}{D1,D}

\fmf{fermion,label=$p/2+q$,label.side=right,width=thick}{q1,q}

\fmfblob{0.2h}{blob}
\fmfv{decor.shape=circle,decor.filled=0,decor.size=0.05h}{q1,D1}
\end{fmfgraph*}
}
\\[2cm]
\nonumber
\parbox{30mm}{
\begin{fmfgraph*}(10,5)
\fmfpen{thick}

\fmfleft{psi}
\fmfright{psi0}

\fmf{fermion,
        label=\noexpand ${{\noexpand \cal Y}_v}_{ij}^{\alpha_0}$,
        label.side=left,
        width=thick}{psi,blob}
\fmf{phantom}{blob0,psi0}

\fmf{phantom,right,tension=0.35,tag=1}{blob,blob0}
\fmf{phantom,tension=0.35,tag=2,left}{blob,blob0}

\fmfposition

\fmfipath{p[]}
\fmfiset{p1}{vpath1(__blob,__blob0)}
\fmfiset{p2}{vpath2(__blob,__blob0)}

\fmfi{dbl_plain,
        label=$p/2 -q$,
        label.side=left,
        width=thick}{
        subpath (0,length(p2)/2) of p2}

\fmfi{fermion,
        label=$p/2 +q$,
        label.side=right,
        width=thin}{
        subpath (0,length(p1)/2) of p1}

\fmffreeze

\fmfforce{point length(p1)/2 of p1 }{q}
\fmfforce{point length(p2)/2 of p2 }{D}

\fmfblob{0.2h}{blob}

\end{fmfgraph*}
}
&=&
\ \parbox{50mm}
{
\begin{fmfgraph*}(10,5)
\fmfpen{thick}

\fmfleft{psi}
\fmfright{psi0}

\fmf{fermion,
        label=\noexpand ${\noexpand \cal X}_v^\beta$,
        label.side=left,
        width=thick}{psi,blob}
\fmf{phantom}{blob0,psi0}

\fmf{phantom,right,tension=0.35,tag=1}{blob,blob0}
\fmf{phantom,tension=0.35,tag=2,left}{blob,blob0}

\fmfposition

\fmfipath{p[]}
\fmfiset{p1}{vpath1(__blob,__blob0)}
\fmfiset{p2}{vpath2(__blob,__blob0)}

\fmfi{fermion,
        label=$ p/2 +k$,
        label.side=left,
        width=thick}{
        subpath (0,length(p2)/2) of p2}

\fmfi{dbl_plain,
        label=$p/2 -k$,
        label.side=right,
        width=thin}{
        subpath (0,length(p1)/2) of p1}

\fmffreeze

\fmfforce{point length(p1)/2 of p1 }{q}
\fmfforce{point length(p2)/2 of p2 }{D}
\fmfforce{point length(p1)/2 of p1 }{q1}
\fmfforce{point length(p2)/2 of p2 }{D1}

\fmffreeze

\fmfshift{(0.2w,0)}{q}
\fmfshift{(0.2w,0)}{D}

\fmffreeze

\fmf{fermion,label=
$-(k+q)$,
 label.side=right,width=thin}{q1,D1}

\fmf{fermion,label=$p/2+q$,label.side=right,width=thin}{q1,q}

\fmf{dbl_plain,label=$ p/2-q$,width=thick,lab.si=left}{D1,D}

\fmfblob{0.2h}{blob}
\fmfv{decor.shape=circle,decor.filled=0,decor.size=0.05h}{q1,D1}
\end{fmfgraph*}}
\ \  +\ \ 
\parbox{50mm}
{
\begin{fmfgraph*}(10,5)
\fmfpen{thick}

\fmfleft{psi}
\fmfright{psi0}

\fmf{fermion,
        label=${{\noexpand \cal Y}_v}_{j'i'}^{\beta_0}$,
        label.side=left,
        width=thick}{psi,blob}
\fmf{phantom}{blob0,psi0}

\fmf{phantom,right,tension=0.35,tag=1}{blob,blob0}
\fmf{phantom,tension=0.35,tag=2,left}{blob,blob0}

\fmfposition

\fmfipath{p[]}
\fmfiset{p1}{vpath1(__blob,__blob0)}
\fmfiset{p2}{vpath2(__blob,__blob0)}

\fmfi{dbl_plain,
        label=$p/2 -k$,
        label.side=right,
        width=thick}{
        subpath (0,length(p1)/2) of p1}

\fmfi{fermion,
        label=$p/2 +k$,
        label.side=left,
        width=thin}{
        subpath (0,length(p2)/2) of p2}

\fmffreeze

\fmfforce{point length(p1)/2 of p1 }{q}
\fmfforce{point length(p2)/2 of p2 }{D}
\fmfforce{point length(p1)/2 of p1 }{q1}
\fmfforce{point length(p2)/2 of p2 }{D1}

\fmffreeze

\fmfshift{(0.2w,0)}{q}
\fmfshift{(0.2w,0)}{D}

\fmffreeze

\fmf{fermion,label=
$-(k+q)$,
 label.side=right,width=thick}{q1,D1}

\fmf{dbl_plain,label=$p/2 - q$, label.side=left,
        width=thick}{D1,D}

\fmf{fermion,label=$p/2+q$,label.side=right,width=thin}{q1,q}

\fmfblob{0.2h}{blob}
\fmfv{decor.shape=circle,decor.filled=0,decor.size=0.05h}{q1,D1}
\end{fmfgraph*}}
\end{eqnarray}
\caption{Graphical representation
of the  Faddeev equation for heavy 
baryons.}
\label{fig-faddeev}
\end{figure}
\end{fmffile}

For the solution of this equation it is convenient to switch to
momentum space. Denoting the total momentum of the baryon 
(with $m_Q v$ subtracted according to the phase factor in
 \gl{trafo}) by $p$
and the relative momentum by $2 q$, $2 k$, respectively, we
obtain (see Fig.~\ref{fig-faddeev}):
\ba
{\cal X}_v^\al(p,q) 
&  = &  
 \frac{i}{2} \, \int_k 
\ D^{\al \al'}(\mi) 
\ g_v(\pl)  
\ \Gamma_v^{\bto} 
\, (\tilde{g}_0(k + q)
\ \Gamma^{\al'})_{ji}
\ {{\cal Y}_v}_{ij}^\bto(p,k)  \ , 
\label{fex} \\
 {{\cal Y}_v}_{ij}^\alo(p,q)  & = &   
\frac{i}{2}  \  \int_k 
\  (D_v)^{\alo \alo'}_{jj'}(\mi) 
\  (g_0)_{ii'}(\pl)  \ (\Gamma^{\bt} \ \tilde{g}_0(k+q))_{i'j'}
\ \Gamma_v^{\alo'} 
\ {\cal X}_v^\bt (p, k) 
\no \\
& + & 
\frac{i}{4} \ \int_k 
\ (D_v)^{\alo \alo'}_{jj'}(\mi) 
\ (g_0)_{ii'} (\pl)
\ \tilde \Gamma_v^{\bto} \ \tilde{g}_v (k + q )
 \ \tilde\Gamma_v^{\alo'}
\ {{\cal Y}_v}^\bto_{j'i'} (p, k) 
\label{fey} \ .
\ea
Here $\pl$ and $\mi$ are abbreviations for $(p/2 + q)$ and 
$(p/2 - q)$, respectively, and $\int_k$ stands for
$\int d^4 k / (2 \pi)^4$.
Note that in momentum space one has
$\tilde g_0(k+q) = C^\dagger g_0^T(-(k+q)) C$ (here, the transpose
 refers to Dirac indices only).
In the following, 
we will restrict ourselves to the lowest lying
diquark states, namely the scalar and axial-vector diquark channel.

Furthermore, we will not perform a full dynamical calculation of the Faddeev
amplitudes, which would require a tremendous amount of numerical work.
Instead, we will consider the diquarks as point-like bosons.
This reduces the Faddeev equation (\ref{fex}), (\ref{fey}) to a 
Bethe--Salpeter type of equation. 
For the light diquarks, we use Klein--Gordon propagators
$D(k^2) = Z [k^2 - M^2]^{-1}$, with masses M  = ${M_s, M_a}$ and
wave function renormalization factors $Z = {Z_s, Z_a}$, where
the indices refer to scalar and axial--vector diquarks, respectively.

Accordingly, for the heavy diquarks we use the propagator of a free bosonic
field  which has been subject to the heavy quark transformation (\ref{trafo}):
\bes
(D_v)^{\bto \bto'}_{jj'}(v\!\cdot\! k) = 
\frac{Z_j}{v\!\cdot\! k - \Delta M_j + i\eps} 
\        \delta^{\bto \bto'} \, \delta_{jj'} 
\quad \times \quad \Big \{ \begin{array}{cc}  1 &  
\mbox{for scalar diquarks} \\   g_{\mu\mu'} - v_\mu v_{\mu'} & 
 \mbox{for axial--vector diquarks} \end{array} \ . 
\ees 
Here, $\Delta M = M - m_Q$ is the residual  diquark mass and
$Z$ denotes the wave function renormalization factor.
The heavy scalar and axial vector diquarks are degenerate
due to the heavy quark spin symmetry.     
The diquark masses and $Z$--factors appearing in the above expressions
are calculated using eqs.~(\ref{ldiq}) and (\ref{hdiq}) (see appendix \ref{B}
for details). Numerical values are given in section~\ref{results}.

The color structure  of the interaction matrices $H$ can 
conveniently be
rewritten in terms of the color projection operators
onto singlet and octet states:  
\bes
H^{A B}_{CE} \ \sim \ 
\frac{i \, \eps^B_{C D}}{\sqrt{2}} \  
\ \frac{i \, \eps^A_{D E}}{\sqrt{2}} \ = \
- {}^{(1)} P_{C E}^{A B} \ + \ \frac{1}{2} \  {}^{(8)} P_{C E }^{A B}
 \ \ , \quad
{}^{(1)} P_{C E}^{A B}  =  \frac{1}{3} \delta^A_C \ \delta^B_E \, . 
\ees
Then eqs.~(\ref{fex}), (\ref{fey}) can be shown to decouple
in color space and the parts describing the relevant
color singlet amplitudes can be identified.
\\
In the case of exact flavor symmetry, the flavor
structure of eqs.~(\ref{fex}), (\ref{fey}) is
trivial, since all propagators are proportional to the unit
matrix
in the (light) flavor space.
Then the flavor matrices $t_{ij}^{a}$
connected with the fields ${\cal X}^a$
project out either the symmetric ($6_F$) or
the anti--symmetric ($\bar 3_F$) flavor part of 
${\cal Y}_{ij}$, respectively, which leads
to a total decoupling of $6_F$ and $\bar 3_F$ amplitudes.\\
The explicit breaking of the $SU(3)_F$ flavor symmetry 
induces a mixing of these amplitudes.
As a first approximation we will keep the flavor structure in the
kernel of eqs.~(\ref{fex}), (\ref{fey}),
 but neglect the mixing effects for the amplitudes.
The error involved is of first order in $(m_s - m_u)$ for the baryon wave
functions, i.e.~of second order for the mass spectrum.\\
\\
Furthermore, we are interested only in ground state baryons, which are
given by s--wave (quark--diquark) amplitudes as discussed in \cite{KKP}.
For s--wave baryons, the transversal momentum of the exchanged quark
 should be of minor importance  and will be neglected.
This approximation implies the replacement:
\be
\sla{k} \longrightarrow (\vk) \, \sla{v} \quad \Rightarrow \quad
\tilde g_0(k) \ \to \ \frac{1}{-\vk  - m + i \epsilon}
\label{transstat} \ ,
\ee
where the relations $\Pp {\cal X}_v = {\cal X}_v$  
and $\Gamma_v \ \Pp = \Pm \ \Gamma_v$
were used. 
Now we insert (\ref{transstat}) into eqs.~(\ref{fex}), (\ref{fey})
and perform the integration over the relative momentum $q$,
defining ${\cal X}(p) = \int_q {\cal X}_v(p,q)$ 
and ${\cal Y}_v(p) = \int_q {\cal Y}_v(p,q)$.
This leads to a decoupling of the terms $\Pm \ {\cal Y}_v$
 and the
 longitudinal parts  $v_\mu {\cal X^\mu}$ (where ${\cal X^\mu}$ denotes an
amplitude containing an axial--vector diquark).
Therefore, the equations now enforce the Bargmann--Wigner
condition, i.e.~to leading order in $1/m_Q$ the baryon amplitudes
are eigenstates of the projector $\Pp$, and the transversality
condition (\ref{transv}) is fulfilled (see \cite{KKP} for more details about
these conditions).

Finally,  linear combinations of the baryon amplitudes 
have to be found
which are eigenstates of projection operators onto spin 1/2 and spin 3/2.
For this purpose, we introduce the following set of spin projection
operators:
\ba
t_{\mu} & = & \frac{1}{\sqrt{3}} (\gamma_{\mu} - \slash{v} v_{\mu}) \ , \quad
v^{\mu} t_{\mu} \ = \ 0 \ , \no \\ 
{}^{\left[3/2\right]} P_{\mu \nu} & = & \ g_{\mu \nu} \ - \ v_{\mu} v_{\nu} \ - \ 
t_{\mu} t_{\nu} \ , \quad v_{\mu} {}^{\left[3/2\right]} P^{\mu \nu}  = 
 \gamma_{\mu} {}^{\left[3/2\right]} P^{\mu \nu} \ = \ 0  \ .
\ea
Here,  ${}^{\left[3/2\right]} P_{\mu \nu}$ projects onto spin~3/2 while
$t_\mu t_\nu$ projects onto the part of spin~1/2 which is transversal
with respect to $v_\mu$.
The Dirac structure in (\ref{fex}), (\ref{fey}) stemming from the
interaction vertices can now be rearranged as\footnote{
Note that since all amplitudes are transversal now, the vertices
$O^\as$ can be brought into a similar form as $O_v^\as$.}
\ba
O^{\as'}_v \, \times \, O^{\as}_v
& = & \left \{ 
\gamma_5 , \ \gamma_{\nu} - v_{\nu} \slash{v}
\right \}^{\as'} \, \times \,
\left \{ \begin{array}{c} 
\gamma_5 \\ \gamma_{\mu} - v_{\mu} \slash{v}
 \end{array} \right \}^{\as}  
 \  = \  
\left \{ \begin{array}{cc} 
1  & \sqrt{3} t_{\nu} \gamma_5 \\
\sqrt{3} \gamma_5 t_{\mu} & 2  \ {}^{\left[3/2\right]} P_{\mu \nu} - t_{\mu} t_{\nu} 
\end{array} \right \}^{\as\as'}  \ .
\label{spinmatrix}
\ea
The system of integral
equations (\ref{fex}), (\ref{fey})
can then be diagonalized with 
respect to the spin structure by introducing 
 the following linear combinations:
\ba
{}^{\left[1/2\right]} {\cal X}_v &  := &   {\cal X}_v  \  , \no \\
{}^{\left[1/2\right]} {\cal X}_v' & := &\frac{1}{\sqrt{2}}
                     \  t_{\mu} \gamma_5 {\cal X}_v^{\mu} \ ,
\no \\
{}^{\left[3/2\right]} {{\cal X}_v^*}^{\mu} & := &\frac{1}{\sqrt{2}}
  \  {}^{\left[3/2\right]} P^{\mu}_{\nu} {\cal X}_v^{\nu} \ .
 \label{projx} 
\ea
and
\ba
{}^{\left[1/2\right]} {\cal Y}_v &  := &  
\frac{1}{2} (  {\cal Y}_v  +  
 \sqrt{3} t_{\nu} \gamma_5 {\cal Y}_v^{\nu} ) \  , 
\no \\
{}^{\left[1/2\right]} {\cal Y}_v' & := & 
\frac{1}{2} ( \sqrt{3} {\cal Y}_v  +  
t_{\nu} \gamma_5 {\cal Y}_v^{\nu}) \ , \no \\
{}^{\left[3/2\right]} {{\cal Y}_v^*}^{\mu} & := &
 {}^{\left[3/2\right]} P^{\mu}_{\nu} {\cal Y}_v^{\nu} \ .
\label{projy}
\ea
Note that in the case of ${}^{\left[1/2\right]} {\cal Y}_v$ 
and ${}^{\left[1/2\right]} {\cal Y}_v'$ both the heavy
scalar and heavy axial vector diquarks contribute. \\
Now eqs.~(\ref{fex}), (\ref{fey}) can be decomposed into 
two separate sets of coupled equations for the baryons
with the light flavors in the $\bar 3_F$ representation
($\Lambda_Q$, $\Xi_Q$) and the $6_F$ representation
($\Sigma_Q^{(*)}$, $\Xi_Q'{}^{(*)}$, $\Omega_Q^{(*)}$),
respectively. The equations for the fields 
${}^{\left[3/2\right]}{{\cal X}_v^*}^a$, 
${}^{\left[3/2\right]}{{\cal Y}_v^*}_{ij}$ indeed turn
out to be identical to those for the fields
${}^{\left[1/2\right]}{{\cal X}_v'}^a$, 
${}^{\left[1/2\right]}{{\cal Y}_v'}_{ij}$ which
reflects the degeneracy due to the heavy quark
spin symmetry.
\\
The fact that the system is now diagonal with respect to the
flavor and the spin content shows that the Pauli principle 
(which was inherent in the amplitude ${\cal X}_v$ containing the light
diquark) is
established dynamically in the amplitude ${\cal Y}_v$.
As shown above, the physical baryon which will be a linear combination of
  ${\cal X}_v$ and  ${\cal Y}_v$, always has
a definite relation between
flavor and spin quantum numbers.
\section{Numerical results}
\label{results}
Even with the simplifications introduced in the previous section,
a complete solution of the resulting coupled integral equations
for the baryon amplitudes ${\cal X}_v(p), {\cal Y}_v(p)$
would be a rather complicated numerical task.\\
To get a first rough estimate of the baryon spectrum, we apply an even
stronger final approximation which has proven to work quite well
in the light baryon sector (compare \cite{BAR92a,BAR92b,HK94}). 
This approximation implies the neglect of the total momentum dependence
of the exchanged quark in $H$, i.e.~$g_0(k) \  \rightarrow \ - 1/m$ 
with $m$ being the light quark mass,
and is referred to as a static approximation.
Here, we will use the same  approximation also for the 
propagator of the exchanged heavy quark,
since, as can be seen from eq.~(\ref{transstat}), it differs from the
light quark propagator only by a constant shift in the relative momentum
of the amount $m v^\mu$. This shift can 
be absorbed into the definition of the wave functions
entering eqs.~(\ref{fex}), (\ref{fey}).

With these  approximations for the exchanged quarks, the 
integral equations (\ref{fex}), (\ref{fey})  for the baryon amplitudes
become a set of completely decoupled algebraic equations
for the masses of spin 1/2 baryons in the $\bar 3_F$  representation
and for those in the $6_F$ representation, the latter including the 
spin 3/2 baryons.
The final equations are of the form
\be
\frac{4} {(F_{ij} + F_{ji} ) } \ - \ {F^{[s]}}_{ij} \  = \ 1   \ . 
\label{fin}
\ee 
where $F_{ij}(vp)$  and $F^{[s]}_{ij}, s = 1/2, \ 3/2$, defined in 
appendix \ref{B}, are integrals over
the relative momentum of the quark and diquark propagators.
These integrals are UV--divergent. In a complete analysis of eqs.~(\ref{fex}),
(\ref{fey}), the integration over the relative momentum $q$ would be limited
to the width of the wave--function appearing in the integral, which
would render the integration UV--finite (compare e.g.\
\cite{IBY93}).
Here, we will employ an UV--regularization for the 
quark--diquark relative
momentum integrals, where for simplicity the same
method will be used as in the calculation of the diquark masses
(see appendix \ref{B}).
 
The usual NJL model parameters 
are fixed in the light
pseudoscalar meson sector and the light baryon sector.
For the latter we refer to the NJL model calculation in \cite{BAR92b}.
In this work, a constituent mass of $m_u = m_d = 450$~MeV
was used, from which the NJL cut--off and the strange quark constituent
 mass are calculated by a fit to the light meson spectrum,
resulting in  $\Lambda = 630$~MeV and $m_s = 650$ MeV.
Moreover, there it was found that 
the coupling constant for the light axial vector diquark has to be
increased relative to that of the scalar sector:  $G_{1a} = 1.5 \
G_1$.
In the following, the value of the coupling 
constant in the heavy sector $G_3$ is left equal to that of $G_1$.
For convenience,  the corresponding
 mesonic coupling constant 
$G_{\rm meson} = 19.4 \ \Lambda^{-2}$ from the
above fit is used as a normalization for $G_1$.

First, some diquark masses for different values
of $G_1$ are presented in Table~\ref{tmass}.
The results for the masses of heavy diquarks 
are similar to those of heavy mesons \cite{EFFR94}.\\
\begin{table}[htb]
\begin{center}
\begin{tabular}{c|| l l | l l l | l l }
$G_1/G_{\rm meson}$ &
$M_s^{\left[u,d\right]}$ & $M_s^{\left[u,s\right]}$ &
$M_a^{\left\{u,d\right\}}$ & $M_a^{\left\{u,s\right\}}$ & 
    $M_a^{\left\{s,s\right\}}$ &
$\Delta M_u$ & $\Delta M_s$ \\
\hline
1.1 & 
705 & 895 &
875 & 1050 & 1215 &
360 & 530 \\
1.2 &
680 & 870 &
865 & 1035 & 1200 &
345 & 520 \\
1.3 &
650 & 845 &
855 & 1020 & 1185 &
330 & 505 \\
1.4 &
620 & 820 & 
845 & 1010 & 1170 &
320 & 495 \\
1.5 &
595 & 795 & 
835 & 1000 & 1160 &
305 & 480 \\
\end{tabular}
\end{center}
\caption{Light scalar and axial vector diquark masses 
and heavy diquark masses (all in MeV) for $m_u=450$~MeV,
$m_s = 650$~MeV, $\Lambda=630$~MeV, $G_3=G_1$, $G_{1a} = 1.5 \, G_1$,
 $G_{\rm meson}=19.4~\Lambda^{-2}$.}
\label{tmass}
\end{table}
\begin{figure}[htb]
\begin{center}
\psfig{file = 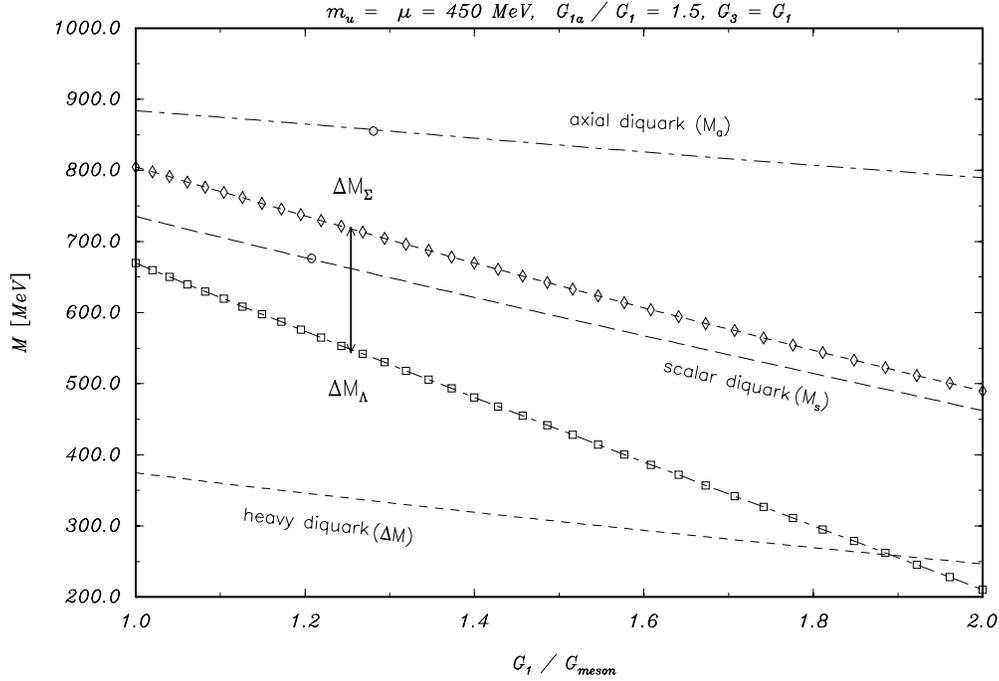, bb = 70 60 520 710, width = 9.5 cm, angle = 90}
\end{center}
\caption{Light diquark masses $M_s$, $M_a$, heavy diquark
residual masses $\Delta M$ and heavy baryon residual masses 
$\Delta M_{\Lambda_Q}$, $\Delta M_{\Sigma_Q}$ as
functions of the coupling strength.
The mass splitting between $\Lambda_c$ and
$\Sigma_c$ found in experiment is
marked with an arrow.}
\label{figure5}
\end{figure}
In Fig.~\ref{figure5},
diquark masses
and residual masses of
the non--strange heavy baryons $\Lambda_Q$ and $\Sigma_Q$
following from eq.~(\ref{fin})
are shown as a function of the
coupling constant $G_1$.
The circles indicate the masses of the diquarks 
found in the fit to the light 
baryons\footnote{In this work a different regularization method was applied,
therefore the circles appear at slightly distinct
values of $G_1$.} \cite{BAR92b}
which determines the parameter range to be used in the following
as $G_1/G_{\rm meson} = 1.2 - 1.3 $.
Indeed, here
we find  a reasonable mass--splitting of 160--175~MeV for the 
heavy baryons $\Lambda_Q$ and $\Sigma_Q$.
The experimental value for the charmed baryons
is 
$M_{\Sigma_c} - M_{\Lambda_c} = 170$~MeV which is
indicated with an arrow in Fig.~\ref{figure5}.\\
The following masses of charmed baryons have been determined so
 far \cite{PD,cleo}:
\ba
\Lambda_c(2285) \ , \quad 
\Sigma_c(2455) \ , \quad
\Xi_c(2470) \ , \quad
\Omega_c(2705) \ , \quad
\Xi_c^*(2644) . 
\no
\ea
In the bottom sector only the data for $\Lambda_b(5641)$ \cite{PD} 
is reliable up to now.\footnote{There is first experimental evidence
for 
$\Sigma_b(5814)$ and $\Sigma_b^*(5870)$ \cite{delphi}.}\\
Our results in Table~\ref{t1} were obtained without changing the parameter
 set\footnote{For comparison, also results for $G_1 = 1.4 \ 
 G_{\rm meson}$ are shown.} 
used for the calculation of light meson and baryon spectra in
{\cite{BAR92b}. 
For a value of $G_1 = 1.3 \ G_{\rm meson}$,
the mass splittings $\Sigma_c  - \Lambda_c$ and 
$\Xi_c - \Lambda_c$ obviously agree quite well with the experimental data.
Note that the (preliminary) result $\Sigma_b - \Lambda_b = 173$ MeV
\cite{delphi} equals the corresponding mass splitting in the charmed sector.
The $\Omega_c - \Lambda_c$ splitting comes out  somewhat too large,
whereas the splitting  $\Xi_c^\ast - \Xi_c$ is somewhat too small. \\
However, our calculation without $1/m_Q$ corrections cannot account
precisely for the mass splittings within the $6_F$ multiplets.
Using spin--weighted averages for these multiplets (with 
preliminary results or estimates \cite{jenkins}
 for the candidates not yet determined)
 would significantly
improve our predictions for $\{\Xi',\Xi^*\}$ and $\{\Omega, \Omega^*\}$
(whereas
the prediction for $\{\Sigma',\Sigma^*\}$ would probably be somewhat too small
then). But a definite analysis requires more reliable experimental data,
and perhaps even a re--analysis of the so far performed
classification of heavy baryon states \cite{falk}.

\begin{table}[htb]
\begin{center}
\begin{tabular}{l|llll}
$G_1/G_{\rm meson}$ & $ \Sigma_Q - \Lambda_Q $
& $\Xi_Q - \Lambda_Q $ & $ \Omega_Q - \Lambda_Q $ 
& $ \Xi^\ast_Q - \Xi_Q $ \\
\hline
        1.20 &  162   &  192  &  495  & 136  \\
        1.30 &  175   &  198  &  513  & 146   \\
        1.40 &  189   &  204  &  531  & 156 \\
\hline
Exp.:&  170 & 185  &  420 & 174 \\
\hline 
\end{tabular}
\end{center}
\caption{Heavy baryon mass splittings in MeV. 
The NJL coupling constant $G_1$,
which triggers the diquark properties, is given in units of the
coupling constant of the (light) pseudoscalar meson sector.}
\label{t1}
\end{table}

\section{Summary and outlook}

In this work we have derived the Faddeev
equations for heavy baryons in the quark--diquark 
picture,
starting from a current--current interaction of the 
NJL--type model, where heavy quarks have been included in the
heavy mass limit.

After extensive use of functional integration
techniques, the baryon propagator has 
been obtained as a series of
quark exchange diagrams, which could be summed up to 
yield the bound state equations for the baryon wave functions.

An essential feature of our approach is that 
two different types of quark--diquark configurations contribute
to the baryon wave functions:\
One, where a heavy quark is coupled
to a light diquark, and the other where a heavy
diquark is coupled to a light quark.
The properties of the heavy diquark are
also fixed by the underlying NJL model
and discussed in some detail, showing a
similar mass pattern as in the case of heavy
mesons. 

For a further analysis of the baryon bound state equations,
we approximated the diquarks as point-like particles, 
which reduces the Faddeev equations to effective Bethe--Salpeter
type of equations.
In an s--wave approximation (that means in our case to neglect the
transversal momentum part of the interaction), the Bargmann--Wigner
condition and the transversality condition for the heavy baryon wave functions
are fulfilled exactly.
Projecting onto color singlet states and defining the proper spin projections,
 we then observe a decoupling of the equations for suitably 
chosen linear combinations of wave functions. 
They describe the predicted multiplets of heavy quark spin
symmetry: the lowest--lying spin 1/2 state (a $\bar 3_F$ multiplet
in the light flavor's $SU(3)_F$) and the spin 1/2 state (a $6_F$
multiplet) which
is degenerate with the spin 3/2 state.

The Pauli principle for the two light
quarks inherent in the light
diquark system is established
dynamically for the baryon wave functions:
We have shown that in the $SU(3)_F$ limit
baryonic states (with definite spin)
where the light quarks are in the $\bar 3_F$--representation
do not mix with those states where the light
quarks belong to the $6_F$ multiplet.
This feature is preserved even if  $SU(3)_F$ breaking is included in
leading order.

For a numerical estimate of the mass spectrum, a static approximation
for the  quark--diquark interaction was found, which treats
both heavy and light quark exchange on an equal  footing
and leads to analytically solvable equations. 
By applying a parameter fit obtained from
the light baryon spectrum,
we then arrived at reasonable masses for
the ground state heavy baryons.

As we have shown in ref.\ \cite{EFKR95},
the quark--diquark picture is also useful
to estimate the coupling of heavy
baryons to pions and weak heavy quark currents,
defining the Isgur--Wise form factors.
The results of the present work are 
to be extended in this direction
in the future.

\section*{Acknowledgements}

T.F.\ would like to thank the theory group of 
Hugo Reinhardt for the warm hospitality during
his stay in T\"ubingen in summer '95.
C.K. acknowledges discussions with the theory group, especially
with R. Friedrich and A. Buck.

\begin{appendix}

\renewcommand{\theequation}{\Alph{section}.\arabic{equation}}
\setcounter{equation}{0}

\section{Deriving the effective baryon lagrangian}
\label{A}
After inserting the functional constants (\ref{einsa})--(\ref{einsd})
into the generating functional, the effective action reads:
\ba
S_{\rm eff} & = &  \int_{x,y}  \bar q \ (i \ \sla{\partial} - m) \ q 
\  +  \ \bar Q_v \ ( i v \cdot \partial ) \ Q_v \  \label{seff} \\
& - & \frac{1}{2} \int_{x,y}  \left( \bar q \ \Delta^{\al} \ 
\Gamma^{\al}  \ q^c \ + \ \bar q^c \ \Delta^{\da \ \al}  \ \Gamma^{\al} \ q
\ + \ \bar Q_v  \Gamma^{\alo}_v  \ \Delta_v^{\alo}
q^c \ + \ \bar q^c \ \Gamma^{\alo}_v \Delta_v^{\da \ \alo} 
Q_v \right) \no \\
& - &  \int_{x}  \left [ \frac{1}{8 G_1} \Delta^{\da \ \al}(x)
\Delta^{\al}(x) \  -  \int_{y} \bar Q_v(y) X^{\al}(y,x) 
\Delta^{\da \ \al}(x) \right ]  \no \\
& +  &  \int_{x,y} \left [ \Delta^{\al}(x) \bar X^{\al}(y,x)
  Q_v(y) 
\ - \ 8 G_1 \int_{y'} \bar Q_v(y')  X^{\al}(y',x) \bar X^{\al}(y,x) Q_v(y)
\right  ] \no \\
& - &  \int_x  \left [  \frac{1}{4 G_3} \  \Delta_v^{\alo}(x)
 \Delta_v^{\da \ \alo}(x) \ -   \int_y \bar q(y) \ t^a \ Y^{\al}(y,x)
 \Delta^{\da \ \alo}_v(x) \right ] \no \\ 
& + &  \ \int_{x,y}  \left [ \Delta^{\alo}_v(x)  \bar Y^{\al}(y,x) \ t^a \ q(y)
\ - \ 4 G_3 \int_{y'} \bar q(y) \ t^a \ Y^{\al}(y,x) \bar Y^{\al}(y',x) 
\ t^a \  q(y') \right ] \no \\
& + &  \int_{x,y}  \bar X \ {\cal X}  \ + \  \bar {\cal X} \ X \ + \ 
 \bar Y \  {\cal Y} \ + \ \bar {\cal Y} \  Y  \no
\ . 
\ea 
The heavy quark fields are removed from the generating functional by a
simple Gaussian integration. For the terms containing light quark fields 
we employ the Nambu-Gorkov formula:
\ba
& & \int D \bar q  \ D  q \ \exp  \ i \left[ \  \frac{1}{2} \int_{x,y}
 ( \bar q ,  \ q^T ) 
\left(  \begin{array}{cc} A & B \\ C & D \end{array} \right )
\left ( \begin{array}{c} q \\ \bar q^T \end{array} \right ) \ \pm \ 
 \frac{1}{2}  \int_x (\bar q \eta \ + \ \bar \eta q) \right]  \no \\
&  = & \exp \left [ \frac{1}{2} {\rm Tr \ log} \left( \begin{array}{cc} A & B
 \\ C & D \end{array}
\right ) \  - \frac{i}{8} (\bar \eta , \  - \eta^T) 
\left ( \begin{array}{cc} A & B \\ C & D \end{array} \right)^{-1}
\left ( \begin{array}{c} \eta  \\ -\bar \eta^T \end{array} \right ) \right]
 \ ,  
\ea
were in our case the components read: 
\ba
A(x,y) & = & ( i \sla{\partial} - m) \delta(x-y) \ - \ 4 G_3
  \ \int_z  t^a \  Y_{a \alo}(x,z) \bar Y_{a' \alo}(y,z) \
t^{a'} \label{aterm} \no \\ 
& + & 
\frac{1}{4} \left[ C \ \Delta_v^\da(y)
\ S_v(y,x) \ \Delta_v(x) C \right ]^T \no \\
B(x,y) & = & - \Delta(x) \delta(x-y) \  C \ ,  \no \\
C(x,y) & = & - C  \ \tilde \Delta(x) \delta(x-y) \ ,  \no  \\
D(x,y) & = & - A^T(x,y) \no \\
\eta(x) & = & 2 \ t^a \ \int_z Y^{\al}(x,z) \
 \Delta_v^{\da \ \alo} \label{eta1}  \no \\
& - &  \left[ \int_{x',z} \bar X^{\al'}(x',z) \Delta^{\al'}(z)
 \ S_v(x',x) \ \Delta_v(x) \ C \right]^T \no \\
\bar \eta(x) & = & 2 \int_z \Delta_v^{\alo}(z) \bar Y^{\al}(x,z) 
\ t^a \label{eta2} \no \\
&  - &   \left [ \int_{y,z} C \  \tilde \Delta_v(x) \ S_v(x,y)
 \ \Delta^{\da \ \al'}(z) X^{\al'}(y,z) \right]^T \ . 
\ea
Here, Tr denotes the trace over all indices, including space-time indices.
We define the inverse of the following matrix
\be
\left( \begin{array}{cc} A & B \\ C & D \end{array} \right )^{-1}
=: \left( \begin{array}{cc} g_{11} & g_{12} \\ g_{21} & g_{22} \end{array} 
\right ) \no
\ee
and the quantity
\be
S_v^{-1}(x,y)  := (i v \cdot \partial) \Pp \delta(x-y)
\ - \ 8 \, G_1 \int_a  \ X^{\al}(x,a) \ \bar X^{\al}(y,a) \ .
\ee
Note that the projection operator $\Pp$ is always present next to a
$Q_v$ or $X_v$ field.

The effective action now reads:
\begin{eqnarray}
i \ S_{\rm eff} &  = &  - \frac{i}{8G_1} \int \ \Delta^{\da \ \al} \Delta^{\al}
\ - \ \frac{i}{4G_3} \int \ \Delta_v^{\alo} \Delta^{\da \ \alo}_v \ + \ 
  Tr \ log \ S_v^{-1} \no \\ 
& + & \frac{1}{2} {\rm Tr \ log} 
\left( \begin{array}{cc} A & B \\ C & D \end{array} \right ) 
\ - \ \frac{i}{8} ( \bar \eta , \ - \eta^T) \left( \begin{array}{cc}
g_{11} & g_{12} \\ g_{21} & g_{22} \end{array} \right) \left( \begin{array}{c}
\eta \\ - \bar \eta^T \end{array} \right ) \no  \\
&  -  & i \ \int_{x,y,a,a'}
\bar X^{\al}(x,a) \Delta^{\al}(a) \  S_v(x,y) \  \Delta^{\da \ \al}(a')
 \ X^{\al}(y,a') \no \\
& + &  \int_{x,y}  \bar X \ {\cal X}  \ + \  \bar {\cal X} \ X \ + \ 
 \bar Y \  {\cal Y} \ + \ \bar {\cal Y} \  Y   \ . \label{effbar}  
\ea  
Expanding this expression up to second order in the
baryon fields, we obtain the entries for the matrix $G_B$
defined in \gl{seffbar} as given by
\ba
&& (G_{B\ XX})^{\bt\bt'}(x,x';\, y,y') = 
\Delta^{\bt}(x')  \  g_v(x,y) \ {\Delta^\da}^{\bt'}(y') \label{gxx}
\no  \\
&& \qquad + \frac{1}{4} \, \int d^4a\, d^4z \ \Delta^{\bt}(x')   
   \ g_v(x,z) \ 
  {\Delta_v}_k (z) \ C \,  g_{22}(z,a)_{kl} \,  C  \
 {\Delta_v^\da}_l(a) \ g_v(a,y) \
 {\Delta^\da}^{\bt'}(y')  \  , 
 \no \\
&& (G_{B\ XY})^{\bt\bto'}_{\ i'j'}(x,x';\, y,y') = 
\frac{1}{2} \, \int d^4 a \ 
\Delta^{\bt}(x') \ 
 g_v(x,a) \  {\Delta_v}_k (a)  \ C \,  g_{21}(a,y)_{ki'}  
\ {\Delta_v^\da}_{j'}^{\bto'}(y') 
\ , 
\label{gxy}  \no \\
&& (G_{B\ YX})^{\bto\bt'}_{ij}(x,x';\, y,y') = 
\frac{1}{2}  \,  \int d^4a 
 \ {\Delta_v}_j^{\bto}(x') \  
g_{12}(x,a)_{il} \,  C  \  {\Delta^\da_v}_l(a)  \ g_v(a,y) \  
{\Delta^\da}^{\bt'}(y') 
\ , 
\label{gyx} \no \\
&& (G_{B\ YY})^{\bto \bto'}_{ij \, i'j'}(x,x';\, y,y') =  
{\Delta_v}_j^{\bto}(x')  \ 
g_{11}(x,y)_{i i'} 
\ {\Delta_v^\da}_{j'}^{\bto'}(y') 
\ , 
\label{gyy}
\ea
where 
\ba
g_{11} & = & (1 - g \ \Delta 
 \ \tilde g \  \Delta^\da)^{-1} g \ ,\label{g11} \no \\
g_{22} & = & - g_{11}^T \no \\
g_{12} & = &  ( 1 - g \ \Delta \ 
          \tilde g \ \Delta^\da )^{-1}  g \ 
\Delta \ \tilde g  \ C^{\da}  \ , \no \\
g_{21} & = & C^{\da} ( 1 - \tilde g \  \Delta^\da \ g \  \Delta)^{-1}\  
\tilde g \  \Delta^\da \ g \ \label{g22}  .
\ea
The expression
 $\tilde g$ is defined as $\tilde g \ = \ C^\da \ g^T \ C $,
where $\left\{\right\}^T$ means 'transposed in all indices'
(including coordinate space).
Note that the Dyson expansions for the terms  $g_{ii}$ yield
quark--diquark contributions up to any order.
\section{Calculation of integrals in the proper--time scheme}
\label{B}
\setcounter{equation}{0}
As far as the integrals (in momentum space)
for the diquark propagators (\ref{ldiq}) and (\ref{hdiq})
 are concerned, the proper-time 
regularization procedure is applied to the 
quark determinant appearing in (\ref{seffbar}), such that
Ward-Identities are respected.
The essential techniques are outlined in \cite{WAR92}.
For the integrals with one heavy and one light particle
in the loop, the basic replacement reads (after Wick-rotation): 
\be \frac{1}{q^2 + m^2} \ = 
\int_0^{\infty} ds \ e^{- s (q^2 + m^2)}
 \ \longrightarrow  
\int_\frac{1}{\Lambda^2}^{\infty} ds \  e^{- s (q^2 + m^2)} \ .
\ee
We define
\ba
I^1(m^2)   &:=& 
  \frac{i}{(2 \pi)^4} \, \int^{reg} \, \frac{d^4 q}{q^2 - m^2 + i
 \eps}
\ = \  \frac{m^2}{16 \pi^2}  \, \Gamma (-1, \frac{m^2}{\Lambda^2})  
\ ,\\[0.2em]
I^2(m^2) &:=&
  - \frac{i}{(2\pi)^4} \, \int^{reg} \,
  \frac{d^4p}{(p^2 - m^2 + i\eps)^2}
\ = \  \frac{1}{16\pi^2} \, 
  \Gamma(0,\frac{ m^2}{\Lambda^2})
\ , \\[0.2em] 
I^3_m(v\!\cdot\!p)   &:=& 
  - \frac{i}{(2 \pi ^4)} \, \int^{reg} \, \frac{d^4 q}
{ (q^2 - m^2 + i \eps) \, (v\!\cdot\! q  + v\!\cdot\! p  + i \epsilon) } 
\no \\
&=& 
   \ \ \frac{\sqrt{\pi}}{16 \pi^2} \, 
   \sqrt{m^2 - (v\!\cdot\!p)^2}  \, \Gamma(-\frac{1}{2},
   \frac{m^2 - (v\!\cdot\!p)^2}{\Lambda^2}) \no \\
&& +  \frac{v\!\cdot\! p }{16 \pi^2} \, \int_0^1 \, 
   \frac{dx}{\sqrt{1-x^2}} \, 
   \Gamma(0, \frac{m^2 - x (v\!\cdot\! p )^2}{\Lambda^2})  \ , \\[0.2em]
I^{4}_m(v\!\cdot\! p )  &:=& 
  - \frac{i}{(2 \pi ^4)} \, \int^{reg} \, \frac{d^4 q}
{ (q^2 - m^2 + i \epsilon) \, (v\!\cdot\! q  + v\!\cdot\! p  + i \epsilon ) } 
 \, \left( -\frac{1}{3} \, \frac{q^2 - (v\!\cdot\! q )^2}{m^2} \right) 
\no \\
&=&
  \ \ \frac{\sqrt{\pi}}{32 \pi^2} \, 
  \frac{ (m^2 - (v\!\cdot\! p )^2)^{\frac{3}{2}}}{m^2} \, 
  \Gamma(-\frac{3}{2}, \frac{m^2 - (v\!\cdot\! p )^2}{\Lambda^2}) \no \\
&&  + 
  \frac{v\!\cdot\! p }{32 \pi^2} \, \int_0^1 \,  \frac{dx}{\sqrt{1 -
x}} 
  \, \left(1 - x \frac{(v\!\cdot\! p )^2}{m^2}\right) \, 
  \Gamma(-1, \frac{m^2 - x (v\!\cdot\! p )^2}{\Lambda^2}) \ .
\ea
Here $\Gamma(\alpha,x) =
\int_x^\infty  dt \ e^{-t} \, t^{\alpha -1} $
denotes the incomplete gamma-function.
\subsection{Diquark polarization tensors}
Starting from the expression (\ref{ldiq}) for the inverse of the light diquark
propagator, in momentum space
the polarization tensor of the scalar diquark
reads:
\ba
\pi_{ij}(k) & = & - \frac{i}{4} \, \int_p \frac{ \tr{\gamma_5 \ 
(\slash{p} - \slash{k} + m_i) \  \gamma_5
\ (\slash{p} + m_j)} }{((k-p)^2 - m_i^2 + i \eps) \ (p^2 - m_j^2 + i \eps)}  
\no \\
& = & \frac{1}{2} \left ( I^1(m_i^2) + I^1(m_j^2) +  
 (k^2 - (m_i - m_j)^2) \, \int_0^1 dx \ I^2 (m_x^2(k^2)) \right )
\ .
\ea
The calculation for the axial vector diquark
with different fermion masses $m_i$, $m_j$
in the loop can be divided into a purely transversal part and
an additional contribution proportional $(m_i-m_j) g^{\mu\nu}$.
The latter reflects the explicit breaking of $SU(3)_F$ for
different quark masses.
\ba
  \pi^{ij}_{\mu\nu}
  &=&
  \frac{i}{8} \, \int^{reg} \frac{d^4p}{(2\pi)^4} \,
  \tr{\gamma_\mu \, \frac{1}{\slash{p} - m_j + i \eps} \,
      \gamma_\nu \, \frac{1}{\slash{p} - \slash{k} - m_i + i \eps}}
\no \\
&=&
(g_{\mu\nu} \, k^2 - k_\mu \, k_\nu) \, \pi_t^b
+ g_{\mu\nu} \Delta^b
\ .
\ea
Note that in order to respect the Ward identities for
the $SU(3)_F$ symmetry (as long as $m_i = m_j$), one has to
identify the gauge invariant part of $\pi_{\mu\nu}$.
The proper-time regularization of
the real part of the light quark determinant automatically
preserves gauge invariance.
We obtain:
\ba
\pi_t^{ij}
&=&
  \frac{1}{16 \pi^2} \,
  \int_0^1 dx \ x \,(1-x) 
    I^2(m_{ij}^2(k^2))
\\
\Delta^{ij}
&=&
\frac{m_i - m_j}{2} \ 
\int_0^1 dx \
(\, (1-x) \, m_j - x \, m_i) \
I^2(m_{ij}^2(k^2))
\ea
with the abbreviation
\be
m_{ij}^2 (k^2) := x m_i^2 + (1-x) m_j^2 - x (1-x) k^2 \ .
\ee
The masses for the light axial vector diquark are presented in
Table \ref{tmass} in the text.

Note that the
polarization tensor for the heavy diquarks is very
similar to those for heavy mesons \cite{EFFR94}.
In momentum space we obtain
(see (\ref{hdiq})):
\ba
\pi_i(v \!\cdot\! k) & = & \frac{i}{2} \int_q 
\frac{ v \!\cdot\! k  -  v \!\cdot\! q  +  m_i}
{((q - k)^2  -  m_i^2 + i \eps) \,  (v \!\cdot\! q  + i \eps )} 
 =  \frac{1}{2}  \, \left(
   I^1(m_i^2) + (v \!\cdot\! p  +  m_i ) \, I^3_{m_i} (v \!\cdot\! p) \right)
\no \ .\\
\ea
The heavy diquark masses following from this can also be found in
Table \ref{tmass} in the text.

\subsection{Baryonic self--energy integrals}

\label{base}

The baryonic self-energy integrals which were used for estimating heavy
baryon masses in (\ref{fin}) are defined as following:
\ba
F_{ij} (v\!\cdot\! p ) 
& := & - \frac{i}{\mu} \int_q \ D_{ij}(\frac{p}{2}-q) \ g_v(\frac{p}{2}+q)
\no \\ 
& = & - \frac{i}{\mu} \int_q \frac{Z^s_{ij}}{(p-q)^2 - (M^s_{ij})^2  + i \eps}
 \ \frac{1}{v\!\cdot\! q  + i \eps}
 \no \\
&=& 
\frac{Z^s_{ij}}{\mu} \ 
   I^3_{M^s_{ij}} (v\!\cdot\! p )  \  ,
\\
F^{[s=1/2]}_{ij}(v\!\cdot\! p ) 
& := & - \frac{i}{\mu} \int_q D_{v \, i}(\frac{p}{2}-q) 
\ g_{j}(\frac{p}{2}+q)
\no \\ 
& = &  - \frac{i}{\mu} \int_q 
  \frac{Z_{v \, i}}{v\!\cdot\! p  - v\!\cdot\! q  - \Delta M_i  + i \eps}
  \   
\frac{v\!\cdot\! q  + m_j} {q^2 - m_j^2 + i \eps} 
\no \\
&  = &
\frac{Z_{v \, i}}{\mu} \  (m_j + v\!\cdot\! p  - \Delta M_i) \ I^3_{m_j}
(v\!\cdot\! p  - \Delta M_i) \ + \ \frac{Z_{v \, i}}{\mu} I^1_{m_j}
\, 
\\
F^{[s=3/2]}_{ij} (v\!\cdot\! p ) & := & 
- \frac{i}{\mu} \int_q D_{ij}^\perp(\frac{p}{2}-q)
 \ g_v(\frac{p}{2}+q) \no \\
& = & -  \frac{i}{\mu} \int_q \frac{Z^a_{ij}}{q^2 - (M^a_{ij})^2 + i \eps}
  \ \frac{1}{v\!\cdot\! p  - v\!\cdot\! q  + i \eps} \,
 \left ( 1 - \frac{1}{3} \, \frac{ q^2 - (v\!\cdot\! q )^2}{(M^a_{ij})^2}
\right ) 
\no \\
& = & 
 \frac{Z^a_{ij}}{\mu} 
 \left ( I^3_{M^a_{ij}}(v\!\cdot\! p ) 
+  I^4_{M^a_{ij}}(v\!\cdot\! p ) \right ) 
\ . 
\ea
\end{appendix}

\end{document}